\documentclass[twocolumn]{article}
\usepackage[utf8]{inputenc}
\usepackage[margin=17mm]{geometry}
\usepackage{fancyhdr}
\usepackage{amsmath}
\usepackage{amsfonts}
\usepackage{amssymb}
\usepackage{graphicx}
\usepackage{verbatim}
\usepackage{siunitx} 
\DeclareSIUnit{\mas}{mas}
\DeclareSIUnit{\arcsec}{as}
\usepackage{pgf-pie}
\usepackage{acronym} 

\usepackage[long,nodayofweek]{datetime}

\usepackage{authblk}
\usepackage{hyperref}

\providecommand{\keywords}[1]
{
  \small	
  \textbf{\textit{Keywords---}} #1
}
\title{Earth rotation parameter estimation from LLR and impact of non-tidal station loading}
\author[1,2]{Vishwa Vijay Singh\footnote{Corresponding author:\newline Vishwa Vijay Singh, email: \href{mailto:singh@ife.uni-hannover.de}{singh@ife.uni-hannover.de}}}
\author[1]{Liliane Biskupek}
\author[1]{Jürgen M\"uller}
\author[1]{Mingyue Zhang}
\affil[1]{\small Institute of Geodesy (IfE),\ Leibniz University Hannover,\ Schneiderberg 50,\ 30167 Hannover,\ Germany}
\affil[2]{\small Institute for Satellite Geodesy and Inertial Sensing, German Aerospace Center (DLR), Callinstraße 36, 30167 Hannover, Germany}

\usepackage{framed}
\usepackage{csquotes}
\usepackage{enumitem}
\usepackage[para,online,flushleft]{threeparttable}
\usepackage{breakcites}
\usepackage{multirow}
\usepackage{natbib}
\setcitestyle{square}
\usepackage{subcaption} 
\usepackage{tabularx}

\usepackage[para,online,flushleft]{threeparttable}
\newcolumntype{P}[1]{>{\centering\arraybackslash}p{#1}} 

\usepackage[multiple]{footmisc} 

\usepackage{url}
\begin{document}

\maketitle
\thispagestyle{fancy}




\begin{abstract}
\ac{LLR} measures the distance between observatories on Earth and retro-reflectors on Moon since 1969. In this paper, we estimate the Earth Rotation Parameters (ERP; terrestrial pole offsets, $x_p$ and $y_p$, and Earth rotation phase, $\Delta$UT) using LLR data. We estimate the values of $\Delta$UT, and the pole offsets separately. For the pole offsets, we estimate the values of $x_p$ and $y_p$ together and separately. Overall, the uncertainties of ERP from the new LLR data (after 2000.0) have significantly improved, staying less than \SI{20}{\micro\second} for $\Delta$UT, less than \SI{2.5}{\mas} for $x_p$, and less than \SI{3}{\mas} for $y_p$ for nights selected from subsets of the LLR time series which have 10 and 15 normal points obtained per night. Furthermore, we add the non-tidal loading effect provided by the \ac{IMLS}, as observation level corrections of the LLR observatories in the analysis. This effect causes deformations of the Earth surface up to the centimetre level. Its addition in the Institute of Geodesy (IfE) \ac{LLR} model, leads to a marginal improvement in the uncertainties (3-$\sigma$ values) of about \SI{1}{\percent} for both, $\Delta$UT and the pole offsets.
\end{abstract}

\keywords{lunar laser ranging;  non-tidal loading; Earth rotation parameters}


\section{Introduction}
\acf{LLR} is the measurement of round trip travel times of short laser pulses between observatories on the Earth and retro-reflectors on the Moon. This has been possible since 1969, when the astronauts of Apollo 11 deployed the first retro-reflector on the lunar surface. Now there are five retro-reflectors on the Moon, and measurements have primarily been carried out from six observatories on the Earth that were or are capable to range to the Moon: the \ac{OCA}, the \ac{MLRS}, the \ac{APOLLO}, the \ac{LURE}, the \ac{MLRO} and the \ac{WLRS}. As the amount of signal loss of the laser pulse is enormous, it is necessary to collect single measurements for 5 to 15 minutes. Of these, a statistically secured mean value is computed, a so called \ac{NP}. Details of the \ac{LLR} measurement process can be found, for example, in \citet{murp13} and \citet{mueller19}. LLR being the longest observation time series of all space geodetic techniques \citep{mueller19} allows the determination of a variety of parameters of the Earth–Moon dynamics, such as the mass of the Earth–Moon system, the lunar orbit and libration parameters \citep{de430_other, pav16}, terrestrial and celestial reference frames and coordinates of observatories and reflectors \citep{mueller2009,hof_etal18} etc. Additionally, it leads to improvements in the solar system ephemerides \citep{kopeikin2008,jplde,pav16}, selenophysics \citep{murp13,hofmann17,viswa19}, and gravitational physics  \citep{will06,mueller_etal12,hof_mu18,Zhang2020,biskupek21}. Furthermore, LLR can also be used to provide tests of Earth orientation parameters \citep{bisk2015,hof_etal18}.


The terrestrial pole offsets (or, \ac{PMC}), $x_p$ and $y_p$, describe the change of the rotation axis in relation to the Earth's surface. The Earth rotation phase $\Delta$UT and the \ac{LOD} refer to the rotational motion of the Earth. All these parameters are summarised as \ac{ERP}. Together with the celestial pole offsets, $\delta X$ and $\delta Y$, as corrections to the conventional precession–nutation model, they define the \acp{EOP}. The \ac{EOP} values are combined from different space geodetic techniques \citep{Bizouard2019}, such as \ac{VLBI}, \ac{GNSS}, \ac{SLR} and \ac{DORIS}, and published\footnote{\url{http://www.iers.org/IERS/EN/DataProducts/EarthOrientationData/eop.html}}\footnote{\url{https://hpiers.obspm.fr/eop-pc/index.php?index=C04&lang=en}} by the Earth Orientation Centre (EOC) of the \ac{IERS}. LLR products are not yet a part of the \acp{EOP} published by the \ac{IERS} (as IERS EOP 14 C04 series). The Kalman Earth Orientation Filter (KEOF) COMB series, of \citet{Ratcliff2020} exists as an alternative to the \ac{IERS} C04 series. It is calculated from the data of \ac{LLR}, \ac{SLR}, \ac{VLBI} and \ac{GNSS}.


The \acp{EOP} obtained from \ac{LLR} and other space geodetic techniques are different in their realisation. \ac{LLR} is the only space geodetic technique which provides a dynamical realisation of the celestial reference system, \citep{zc09}, whereas other space geodetic techniques provide a kinematic realisation of the celestial reference system. Additionally, $\Delta$UT values can only be obtained from \ac{VLBI} and \ac{LLR}, making results from LLR useful in validating the results from VLBI.

In this study, we estimate the Earth rotation phase ($\Delta$UT) and terrestrial pole offsets ($x_p$ and $y_p$) using LLR data. Section \ref{sec:data_meth} defines the data and methods of calculations used for the estimation of $\Delta$UT (in section \ref{sec:dUT}), and $x_p$ and $y_p$ (in section \ref{sec:xpyp}). In section \ref{sec:dUT} and section \ref{sec:xpyp}, we give the most recent results and discuss the accuracy of the \acp{ERP}, analyse the correlations of the estimated parameters with each other, and additionally study the effect of \ac{NTL} on the estimated values of $x_p$ and $y_p$, and $\Delta$UT. In section \ref{sec:conclusions}, we give the conclusions to this study.

\section{Data and method}\label{sec:data_meth}

In Germany, from the early 80ies, the software package \acs{LUNAR} (\acl{LUNAR}) has been developed to study the Earth-Moon system and to determine several related model parameters \citep{Egger1985,Gleixner1986,Bauer1989,Muller1991}. The analysis model used in LUNAR is based on Einstein’s theory of relativity. It is fully relativistic and complete up to the first post-Newtonian ($1/c^2$) level. To take advantage of the high-precision \acp{NP} that can be obtained with an accuracy of several millimetres \citep{murp13}, the \acs{LUNAR} software was updated continuously \citep{bisk2015,hofmann17}. A recent overview of \acs{LUNAR} is given in \citet{hof_etal18}, a detailed description can be found in \citet{Muller2014a}. The adjustment is done in a Gauss-Markov model (GMM) where, in the current version of LUNAR, up to 200 unknown parameters can be determined with their uncertainties.

\subsection{LLR data}\label{sec:llr_data}

\begin{figure}[ht]
    \def\svgwidth{\columnwidth}
    \input{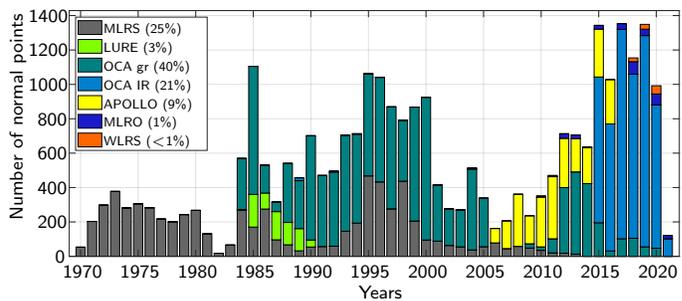}
    \caption{Distribution of the \num{28093} normal points over the time span April 1970 - April 2021. In the legend the percentages of the contribution of the respective observatories are given. The three observatories McDonald, MLRS1 and MLRS2 are linked in the analysis and listed here as \enquote*{MLRS}. OCA measurements with laser wavelength of $\lambda = \SI{684.3}{\nano\metre}$ and $\lambda = \SI{532}{\nano\metre}$ are listed as \enquote*{OCA gr}.}
	\label{fig:distributionNP}
\end{figure}

The distribution of \ac{LLR} \acp{NP} has a big impact on the determination of various parameters. Non-uniform data distribution is one of the reasons for correlations between solution parameters \citep{Williams2009b}. From 2015, many \acp{NP} were measured using laser pulses of infra-red wavelength, due to which ranging near new and full Moon became possible \citep{Chabe2020} for \ac{OCA} and \ac{WLRS}. This leads to a better coverage of the lunar orbit over the synodic month, i.e. the time span in which the Sun, the Earth, and the Moon return to a similar constellation again. With a better coverage of the lunar orbit, it is possible to perform a more uniform estimation of \acp{ERP}. Fig. \ref{fig:distributionNP} shows the temporal distribution of the measured \acp{NP} since 1970. In the analysis, the three observatories McDonald, MLRS1, and MLRS2 are linked by local ties and analysed together, and are therefore listed in the figure as MLRS. OCA measurements with laser wavelength of $\lambda = \SI{684.3}{\nano\metre}$ and $\lambda = \SI{532}{\nano\metre}$ are listed in the figure as OCA green. As given in the legend of Fig. \ref{fig:distributionNP}, more than \SI{60}{\percent} of the \acp{NP} were observed by \ac{OCA} (\SI{40}{\percent} with green and \SI{21}{\percent} with infra-red (IR) laser light). In the last years, only \ac{OCA} and \ac{APOLLO} provided regular \acp{NP}, and \ac{MLRO} and \ac{WLRS} only contributed a few \acp{NP}. For the year 2020, \SI{84}{\percent} of the \acp{NP} were measured by \ac{OCA} in IR. Therefore, with this distribution of \acp{NP}, the overall analysis is dominated by the \ac{OCA} \acp{NP}. 

The measured \acp{NP} serve as observations in the analysis. They are treated as uncorrelated for the stochastic model of the least squares adjustment and are weighted according to their measurement accuracies.

\subsection{ERP a-priori data and estimation}\label{subsec:erp_data_est}

In LUNAR, as a-priori \acs{ERP} values, the KEOF COMB2019 series \citep{Ratcliff2020} is used until 0h UTC 01.01.1983 and the IERS EOP 14C04 series is used from 0h UTC 02.01.1983 onward. The differences between KEOF COMB2019 and IERS 14 C04 series are of a few \si{\micro\arcsec} for the \ac{PMC} and a few \si{\micro\second} for $\Delta$UT after January 1, 2000. Before 2000, these differences are up to a few \si{\milli\second} for $\Delta$UT and a few \si{\mas} for the \ac{PMC}. Further details about the differences between IERS 14 C04 series and the COMB series can be found in \citep{Ratcliff2020}. This implementation of two different EOP time series within LUNAR is done as the COMB2019 series includes LLR data in its formation, and therefore fits the initial phase of the observations better. After 01.01.1983, the differences between the two series become smaller, and therefore we use the IERS 14 C04 to benefit from its shorter latency period.

As the rotation matrix between the Earth fixed \ac{ITRS} and the space fixed \ac{GCRS} includes the PMC and $\Delta$UT in its calculation, these can be estimated from LLR analysis, as shown by \citet{dickeyetal1985,Muller1991,bisk2015}. The most recent results of ERP from LLR have been discussed by \citet{hof_etal18}. \citet{bisk2015} gave the equations for the partial derivatives of ERP (used for the adjustment) and discussed the results of the different possible methods to obtain ERP from LUNAR (selecting time spans or specific nights for which ERP are adjusted) for \num{20047} NPs (1970 - 2013), where the accuracy of $\Delta$UT achieved was a maximum of \SI{0.14}{\milli\second} (70 day period of estimation), and the accuracy of the \ac{PMC} was in the range of \num{30} - \num{90} \si{\mas} for different combinations of observatories from which the PMC were estimated ($x_p$ and $y_p$ were estimated individually, due to high correlations with each other). \citet{hof_etal18} discussed the results of estimation of the Earth rotation phase for \num{23261} NPs (1970 - 2016), achieving an accuracy under \SI{0.089}{\milli\second} when estimating $\Delta$UT from all observatories, and under \SI{0.044}{\milli\second} when estimating $\Delta$UT from only OCA and APOLLO.

The ERP estimation from other techniques leads to better results than those from LLR, due to the availability of much more data and a better global coverage. The accuracies achieved from different space geodetic techniques are given in Table \ref{tab:erp_others}, where it can be seen that the best results of the PMC are from GNSS, primarily due to dense network of its stations. The sensitivities of the space geodetic techniques are different for various \acp{EOP}, and not all techniques can determine all parameters, for example, $\Delta$UT values can only be obtained from \ac{VLBI} and \ac{LLR}.

\begin{table}[!htp]
\caption{Accuracies of the ERPs obtained from different space geodetic techniques \citep{Sciarretta2010,SCHUH201268,capitain_eop17,Zajdel2020}.}\label{tab:erp_others}
\centering
\begin{tabular}{cccc}
\hline\noalign{\smallskip}
Technique &\multicolumn{2}{c}{Parameters} \\
&PMC & $\Delta$UT \\
\noalign{\smallskip}\hline\noalign{\smallskip}
VLBI & 50 - 80 \si{\micro\arcsec} & 1 - 5 \si{\micro\second}\\
\noalign{\smallskip}\hline\noalign{\smallskip}
SLR & 10 - 30 \si{\micro\arcsec} & - \\
\noalign{\smallskip}\hline\noalign{\smallskip}
GPS & 5 - 20 \si{\micro\arcsec} & - \\
\noalign{\smallskip}\hline
\end{tabular}
\end{table}

Since 2015, IR \acp{NP} measurements are available from \ac{OCA}. These enable a better coverage of the lunar orbit and obtain more \acp{NP} per night, leading to a better and more stable estimation of \acp{ERP} from LLR, which achieves higher accuracy compared to previous results. For \acp{ERP} determination from the LLR analysis, the whole dataset of \acp{NP} is pre-analysed. Here, different configurations can be taken into account, which also define the configuration in the analysis, thereby making it possible to vary the number of \acp{NP} per night and to estimate \acp{ERP} from the data of all observatories or only for a single observatory.

As shown in Fig. \ref{fig:distributionNP}, the number of \acp{NP} measured per year has significantly risen over the past few years, implying that more \acp{NP} were recorded per night, and that they were recorded for more nights. Despite having more \acp{NP} per night over the past years, it is currently still difficult to estimate \ac{PMC} and $\Delta$UT together for one night and one observatory due to their high correlation with each other. Therefore, in the current study, either \ac{PMC} ($x_p$ and $y_p$ individually, and together) or $\Delta$UT were determined, where the other values were fixed to the a-priori \ac{EOP} series. It is also difficult to determine \acp{ERP} with the coordinates and velocities of the observatories together for one night, therefore, the velocities of the observatories were fixed to the ITRF2014 solution values\footnote{as APOLLO observatory is not included in the ITRF2014 solution, we used the velocity for White Sands GNSS observatory (DOMES number: 49884S001) as the replacement}\footnote{\url{https://itrf.ign.fr/ITRF_solutions/2014/more_ITRF2014.php}}.

The selection of nights was based on the observatory/observatories where the \acp{NP} were measured from (only from APOLLO, only from OCA, or from all observatories). We selected only subsets of nights in which at least 5, 10, and 15 \acp{NP} were measured. Table \ref{tab:subsets_nights} gives details of the subsets which were created for this study, and the abbreviations we use to refer to these subsets.


\begin{table}[!ht]
\caption{Details of the data subsets which were created for this study along with their time spans and abbreviations used for them.}\label{tab:subsets_nights}
\centering
\begin{threeparttable}
\begin{tabular}{cP{1.0cm}P{0.8cm}P{1.9cm}P{1.5cm}}
\noalign{\smallskip}\hline\noalign{\smallskip}
Obs.$^*$ & NPs per night & No. of nights & Abbreviation used & Time span\\
\noalign{\smallskip}\hline\noalign{\smallskip}
\multirow{2}[18]{*}{APOLLO} &5 & 261 & Apollo05 & 04.06.2006 - 26.10.2016 \\ \noalign{\smallskip}\cline{2-5} \noalign{\smallskip}
&10 & 63 & Apollo10 & 29.03.2008 - 24.10.2016 \\ \noalign{\smallskip}\hline \noalign{\bigskip}
\multirow{3}[28]{*}{OCA} &5 & 1320 & OCA05 & 07.04.1984 - 30.03.2021 \\ \noalign{\smallskip}\cline{2-5} \noalign{\smallskip}
&10 & 714 & OCA10 & 08.04.1984 - 24.03.2021 \\ \noalign{\smallskip}\cline{2-5} \noalign{\smallskip}
&15 & 355 & OCA15 & 12.04.1984 - 26.12.2020 \\ \noalign{\smallskip} \hline \noalign{\bigskip}
\multirow{3}[28]{*}{All} &5 & 1975 & All05 & 15.05.1975 - 30.03.2021 \\ \noalign{\smallskip}\cline{2-5} \noalign{\smallskip}
&10 & 914 & All10 & 29.09.1983 - 24.03.2021 \\ \noalign{\smallskip}\cline{2-5} \noalign{\smallskip}
&15 & 450 & All15 & 08.04.1984 - 19.01.2021 \\
\noalign{\smallskip}\hline
\end{tabular}
\begin{tablenotes}
 \item $^*$Observatory/Observatories \\
\end{tablenotes}
\end{threeparttable}%
\end{table}

Preliminary studies showed that the accuracies of the \acp{ERP} determined in our LLR adjustment were too optimistic. Comparisons with other parameters led to the conclusion that a 3$\sigma$-accuracy is a realistic specification for the accuracies in our LLR analysis. For this reason, all accuracies for \acp{ERP} are given as 3$\sigma$-accuracies in the following sections. 


\subsection{Non-tidal loading data}\label{subsec:ntsl}
The IERS 2010 conventions \citep{Petit2010} do not recommend the addition of \ac{NTL} deformations in the calculation of the displacement of a reference point, as their modelling accuracy is low and their impact on the geodetic parameters compared to other deformations is small. However, their inclusion can be beneficial in geodetic analyses, as pointed out in recent studies \citep{glomsda_etal2020,singh_etal_2021}, as the accuracy of the loading effect due to \ac{NTL} has improved over the past years due to the improved accuracy of the numerical weather models used for its calculation \citep{mpiom,gfz_ntsl,merra2,era5}.

The IERS established the Global Geophysical Fluids Center (GGFC) in 1998, which has different bureaus responsible for research and data provision related to the redistribution of masses in atmosphere, oceans, and hydrological (land water) systems. The time series of \acp{NTL} over different time spans, based on calculations using numerical weather models and Green's functions \citep{farrell72,gfz_ntsl,imls_ntsl}, can be obtained from these bureaus. These loadings can be added as observation level corrections in the calculation of the instantaneous position of a reference site.


In this study, we use the \ac{NTL} provided by the International Mass Loading Service (IMLS)\footnote{\url{http://massloading.net/}}. It provides three non-tidal loadings (atmosphere: NTAL, ocean: NTOL, and hydrological: HYDL), which occur due to redistribution of masses in atmospehere, oceans, and land water. For further details we refer the reader to our previous study \citep{singh_etal_2021}, where we discussed the effect of \ac{NTL} as observation level corrections in LUNAR, and gave the details of the differences of the \ac{NTL} data from different data centres, described the different components of the NTLs, including discussions on specifics such as resolution, numerical weather models used for each loading, etc. Overall, from that study, we concluded that the addition of \ac{NTL} in LUNAR, primarily from the IMLS, leads to small benefits in reduction of the weighted root mean square (WRMS) of LLR residuals, reduced the annual signal visible in the LLR residuals, and improved the estimation of station coordinates.


The loading deformation can be defined in a centre of figure (CF) frame (realised from the positions of geodetic stations on the solid Earth) or centre of mass (CM) frame (centre of orbiting satellites). This is controlled by the choice of degree-one load Love numbers, which enter the Green’s function summation \citep{petrov_boy04,gfz_ntsl,imls_ntsl} to calculate the effect of \ac{NTL}. For further details on the differences between CM and CF, we refer to \citet{sun_yu_gcm}. In our LLR analysis, the a-priori station positions are aligned to the CM frame and therefore the loadings used in this study were chosen in the CM frame.

\section{Estimation of \texorpdfstring{$\Delta$}{}UT}\label{sec:dUT}
As mentioned in \ref{subsec:erp_data_est}, the nights for which the \ac{ERP} are estimated are categorised into different subsets. The $\Delta$UT values were estimated for the nights of these subsets, using a-priori values from the combination of KEOF COMB2019 and IERS 14 C04 series, as described above. In the subsections that follow, we discuss the results of the estimated values of $\Delta$UT and their accuracy, their correlations with various parameters, and the effect of \ac{NTL} on $\Delta$UT.

For the estimation of $\Delta$UT, the velocities of the LLR observatories are fixed to ITRF 2014 solution values, and the values of the \ac{PMC} ($x_p$ and $y_p$) are fixed to their a-priori values.

\subsection{Estimated values}\label{subsec:res_dut}
For the estimation of $\Delta$UT, the accuracy of the estimated values becomes better (i.e. smaller standard deviations) over the years. This is due to the improved accuracy of the NPs measured over the years, primarily due to the improved accuracy of the OCA green laser data, the highly accurate IR NP data (starting 2015), and the highly accurate APOLLO data (starting 2006). The differences of the estimated values from LUNAR to the values from IERS 14 C04 also become smaller over the time span, indicating the close agreement of the $\Delta$UT values from the different space geodetic techniques (VLBI and LLR) with highly accurate data, thereby also validating each other.

\begin{figure*}[!ht]
    \begin{subfigure}{.48\textwidth}
    \centering
    \includegraphics[width=1\textwidth]{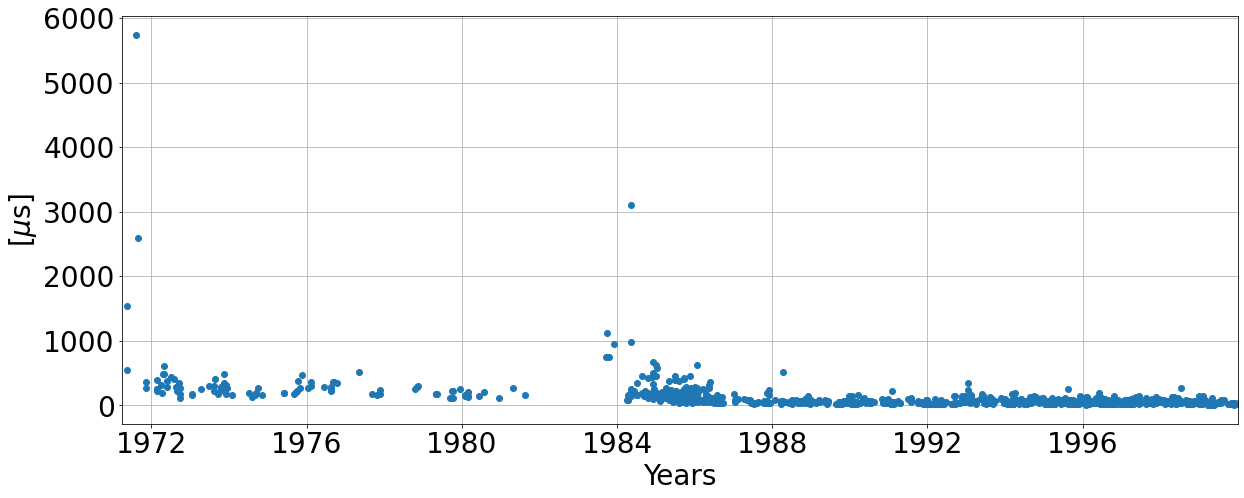}
    \caption{All05}
    \end{subfigure}%
    \qquad
    \begin{subfigure}{.48\textwidth}
    \centering
    \includegraphics[width=1\textwidth]{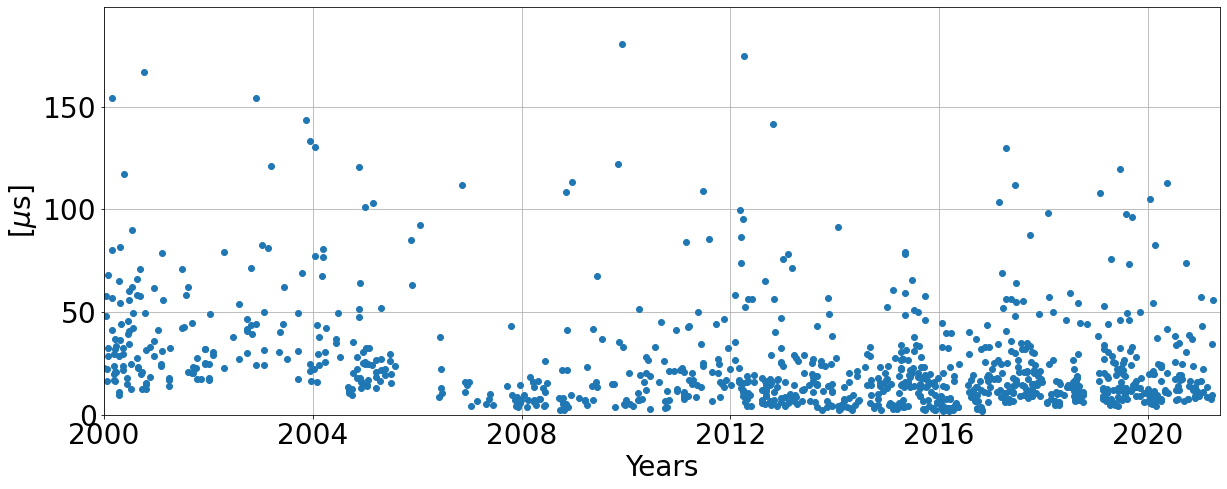}
    \caption{All05}
    \end{subfigure}
    
    \begin{subfigure}{.48\textwidth}
    \centering
    \includegraphics[width=1\textwidth]{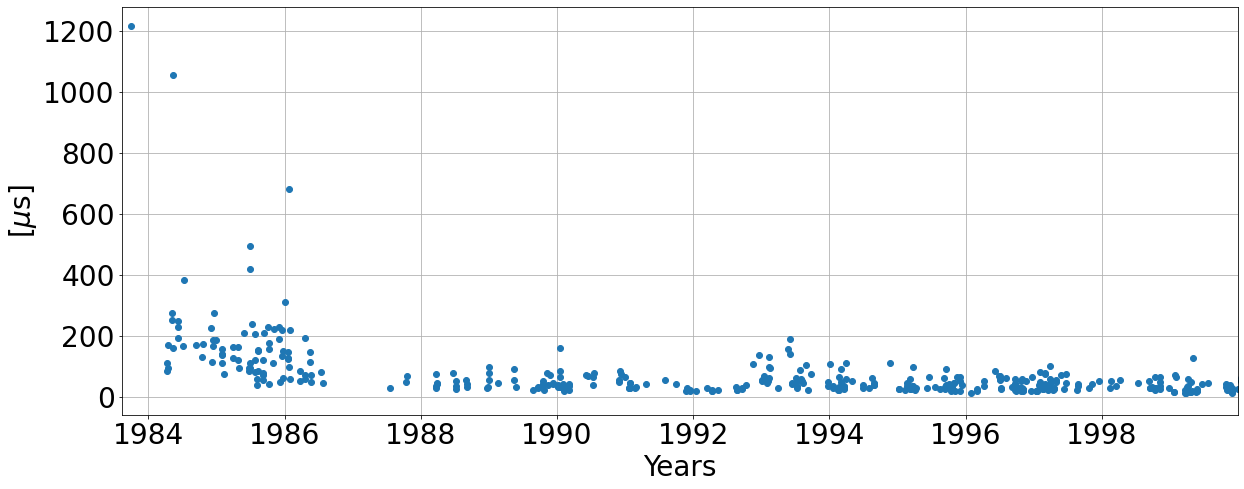}
    \caption{All10}
    \end{subfigure}%
    \qquad
    \begin{subfigure}{.48\textwidth}
    \centering
    \includegraphics[width=1\textwidth]{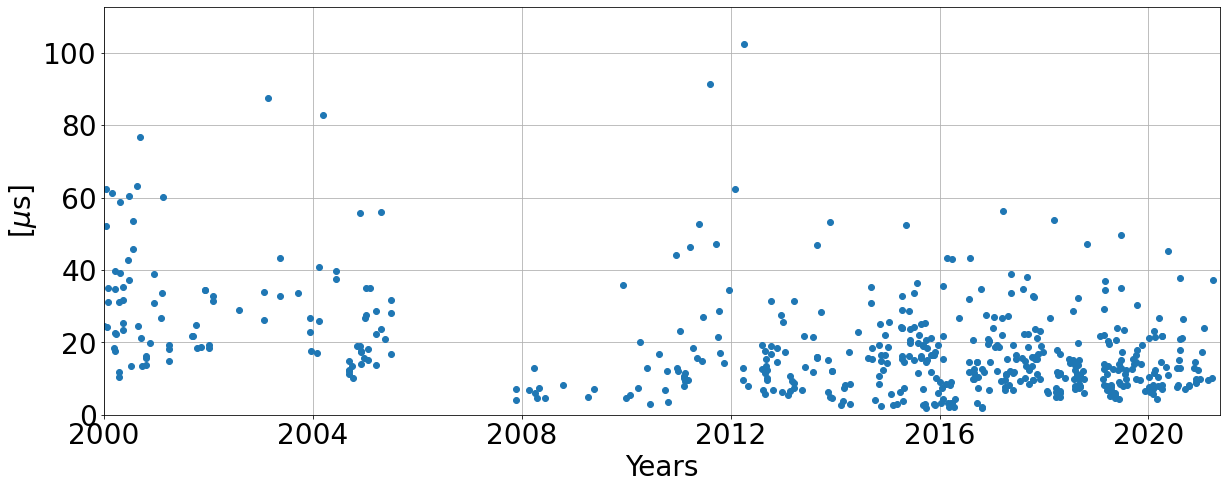}
    \caption{All10}
    \end{subfigure}
    
    \begin{subfigure}{.48\textwidth}
    \centering
    \includegraphics[width=1\textwidth]{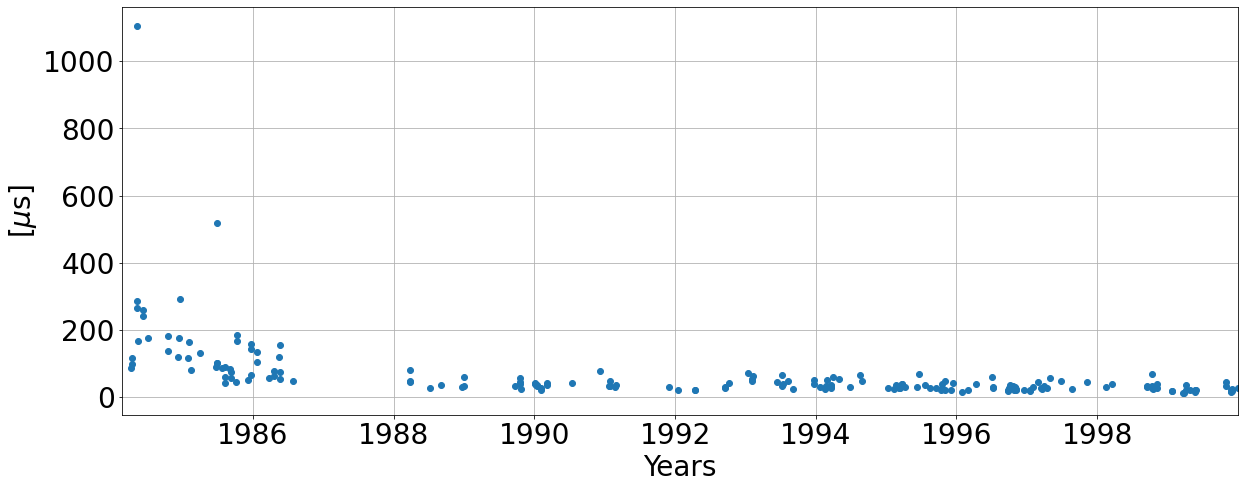}
    \caption{All15}
    \end{subfigure}%
    \qquad
    \begin{subfigure}{.48\textwidth}
    \centering
    \includegraphics[width=1\textwidth]{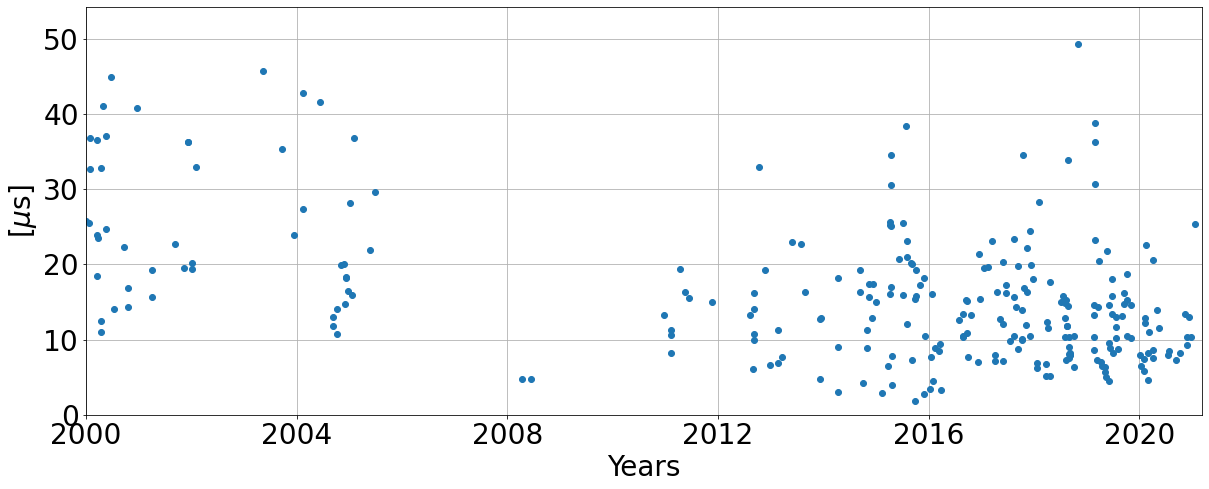}
    \caption{All15}
    \end{subfigure}
    \caption{Accuracy of the estimated values of $\Delta$UT for the subset All05, All10, and All15: in (a), (c), and (e) from the beginning of the dataset until 2000.0, and in (b), (d), and (f) from 2000.0 until the end of the dataset. Note the differences in the range of the axes for each sub-figure.}
    \label{fig:all_dut}
\end{figure*}

\begin{figure*}[!ht]
    \begin{subfigure}{.48\textwidth}
    \centering
    \includegraphics[width=1\textwidth]{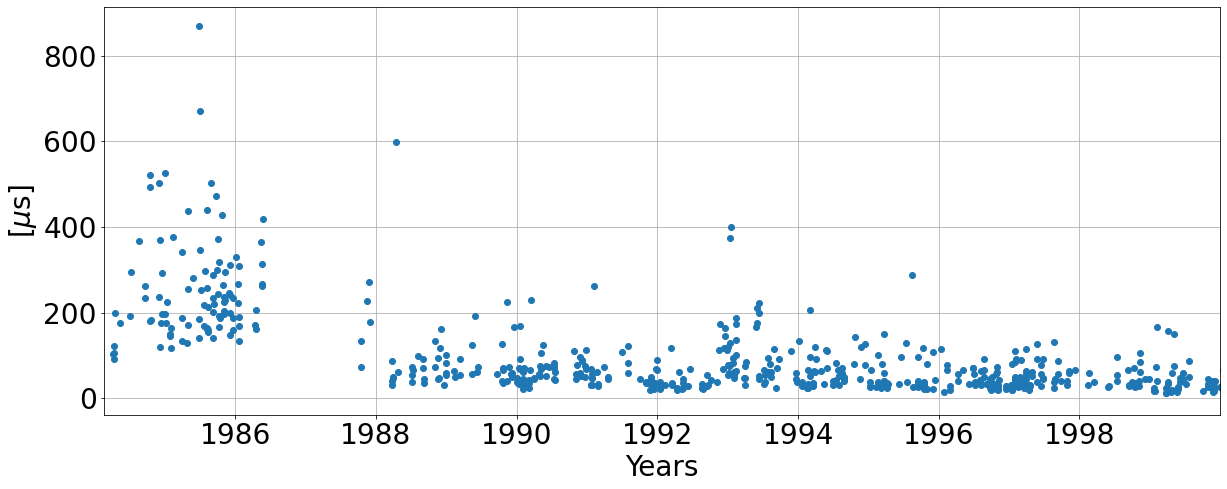}
    \caption{OCA05}
    \end{subfigure}%
    \qquad
    \begin{subfigure}{.48\textwidth}
    \centering
    \includegraphics[width=1\textwidth]{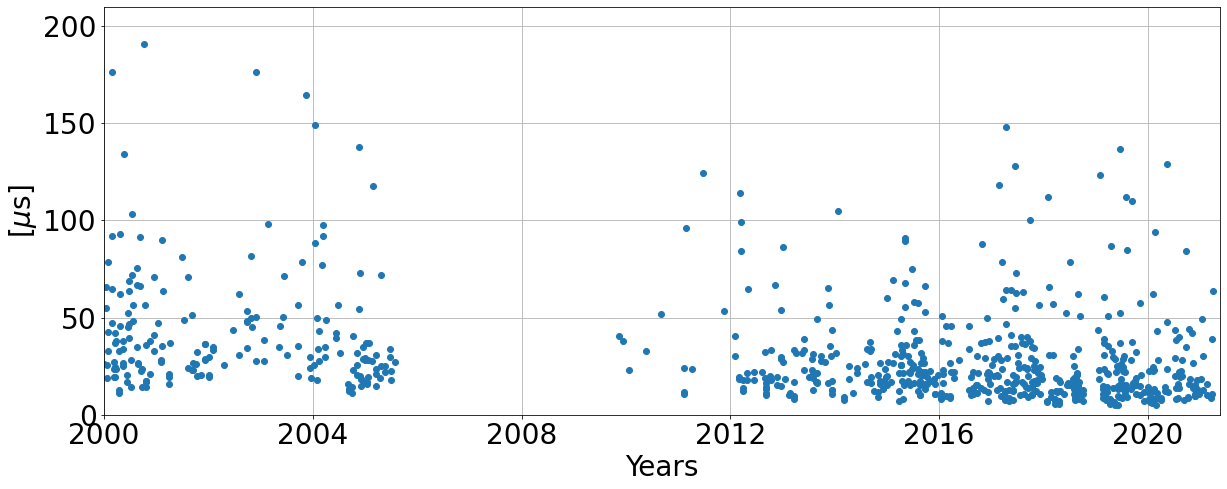}
    \caption{OCA05}
    \end{subfigure}
    
    \begin{subfigure}{.48\textwidth}
    \centering
    \includegraphics[width=1\textwidth]{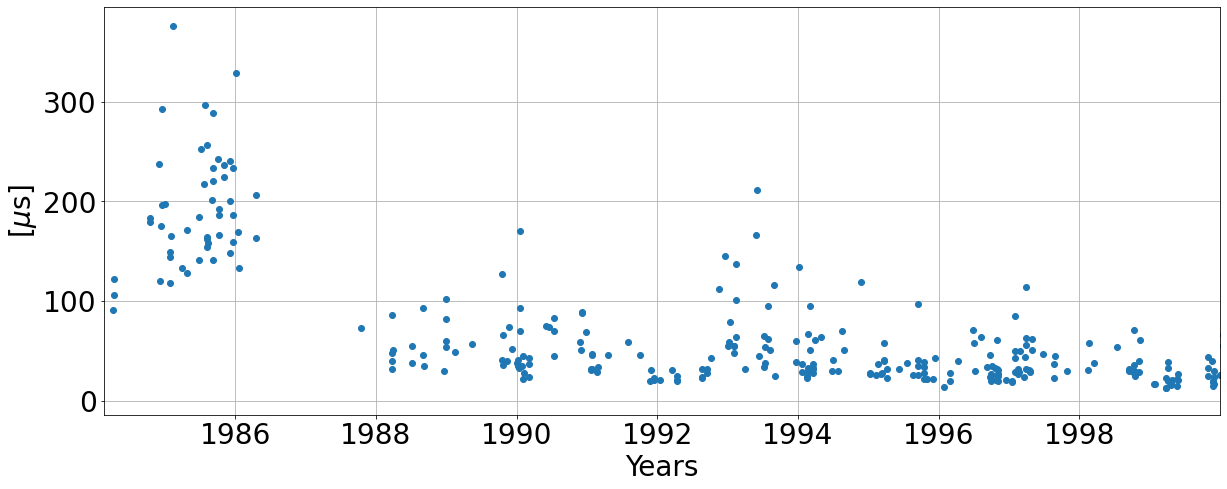}
    \caption{OCA10}
    \end{subfigure}%
    \qquad
    \begin{subfigure}{.48\textwidth}
    \centering
    \includegraphics[width=1\textwidth]{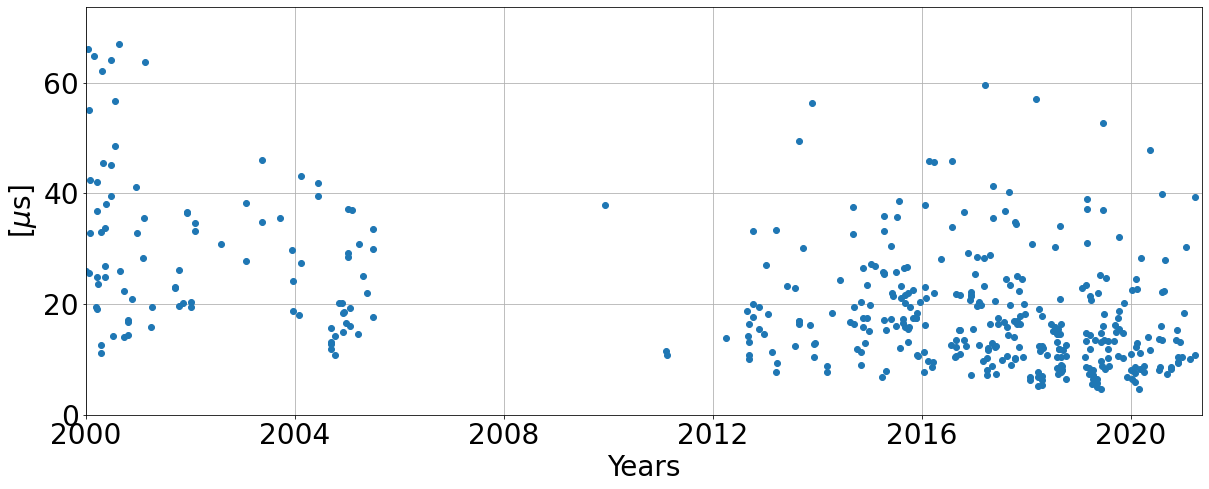}
    \caption{OCA10}
    \end{subfigure}
    
    \begin{subfigure}{.48\textwidth}
    \centering
    \includegraphics[width=1\textwidth]{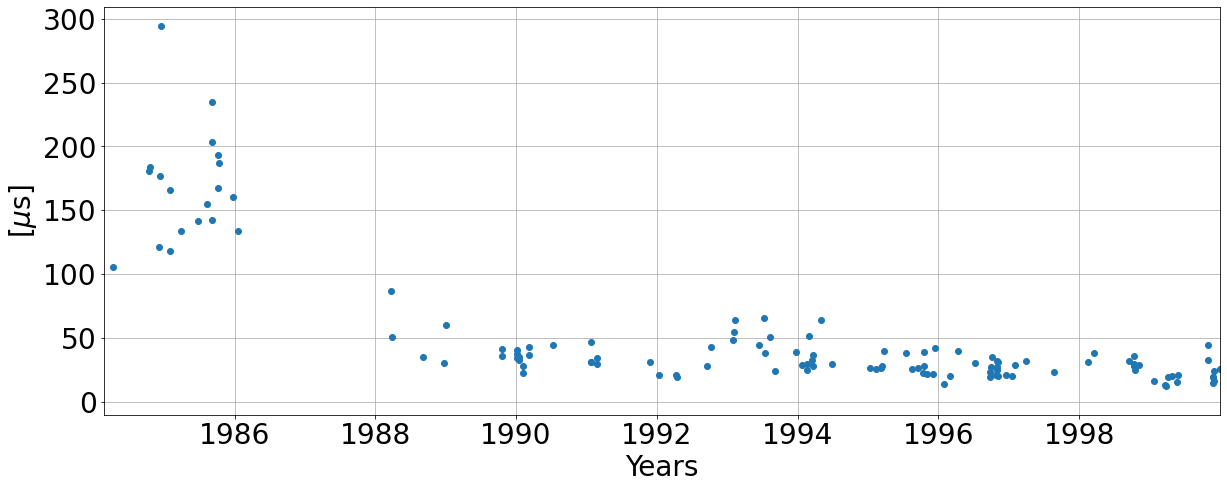}
    \caption{OCA15}
    \end{subfigure}%
    \qquad
    \begin{subfigure}{.48\textwidth}
    \centering
    \includegraphics[width=1\textwidth]{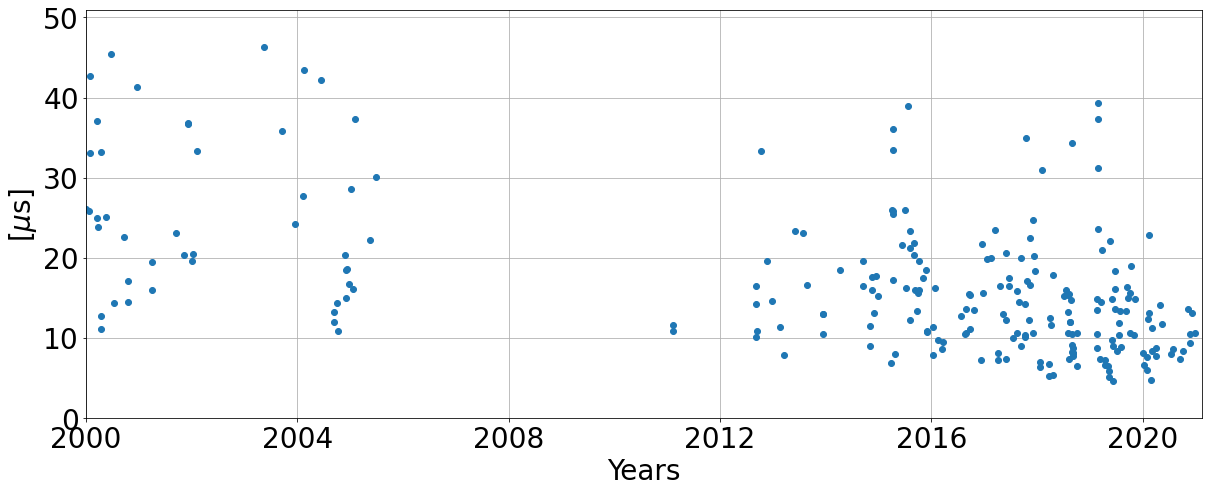}
    \caption{OCA15}
    \end{subfigure}
    \caption{Accuracy of the estimated values of $\Delta$UT for the subset OCA05, OCA10, and OCA15: in (a), (c), and (e) from the beginning of the dataset until 2000.0, and in (b), (d), and (f) from 2000.0 until the end of the dataset. Note the differences in the range of the axes for each sub-figure.}
    \label{fig:oca_dut}
\end{figure*}

Fig. \ref{fig:all_dut} shows the accuracy of the $\Delta$UT values for the subsets All05, All10, and All15, and Fig. \ref{fig:oca_dut} shows the accuracy of the $\Delta$UT values for the subsets OCA05, OCA10, and OCA15. As the accuracy of the $\Delta$UT values show a significant improvement in the recent years, we split the figure of each subset into two time spans, setting a break at 0h UTC 01.01.2000 (henceforth, \enquote*{2000.0}). Additionally, Table \ref{tab:dut_subsets} shows the mean values of all subsets considered in this study and the mean of the differences of the estimated values to the values of the IERS 14 C04 series. The values given in Table \ref{tab:dut_subsets} are also split into two time spans, to show the differences in the estimation from the overall time span to the new data and to stress upon the best possible quality of the estimation from LLR.

The differences to the IERS 14 C04 series are primarily due to the different space geodetic techniques used to obtain the $\Delta$UT values. Moreover, the C04 series is a combined product of different space geodetic techniques, and this combination also contributes to the improvement of the final product. As the results presented in this study are only from \ac{LLR} analysis, a part of the differences could occur due the difference in obtaining the final $\Delta$UT values. A part of these differences are presumed to occur because of the different realisations of the \ac{ICRS} from \ac{VLBI} and \ac{LLR}. Additionally, they could be due to systematic errors in our calculation.

From Fig. \ref{fig:all_dut}, Fig. \ref{fig:oca_dut}, and Table \ref{tab:dut_subsets}, it can be seen that the stricter the selection criteria, i.e. only those nights selected for ERP estimation with a higher number of NPs per night, the better the results. However, it must also be noted that the improvement in the results with the strict selection criteria is more stark in the results until 2000.0, and leads to less improvement in the results after 2000.0, indicating that the very strict selection criteria is needed for data with low accuracy and that with the improvement of the accuracy of the NPs, a less strict selection criteria of the nights can be chosen. Additionally, it can be seen from Table \ref{tab:dut_subsets} that the solution from the subsets of nights chosen from all observatories perform worse than the solution from the subsets of nights chosen from OCA only for the results before 2000.0. This is due to the reason that the accuracy of the NPs involved in the respective time spans of the subsets is significantly better from OCA compared to when NPs from all observatories are chosen. However, when considering the results of all solutions after 2000.0, the results from the solutions of the \enquote*{All} subsets perform better than those from the solutions of the \enquote*{OCA} subsets, as the accuracy of the NPs (in this time span) from all observatories is comparable and the solutions of \enquote*{All} subsets benefit from having more data and having a global coverage of the NPs involved and including the very good \ac{APOLLO} data.

Using the radius of Earth (at the equator) as \SI{6378}{\kilo\meter}, \SI{10}{\micro\second} corresponds to \SI{4.6}{\milli\metre} on the Earth's radius, implying that the best possible (and current) accuracy of estimation of $\Delta$UT (subset All15, after 2000.0) from LLR corresponds to a resolution of \SI{7.3}{\milli\metre} on Earth's surface. Compared to the resolution obtained from VLBI of \SI{0.5}{\milli\metre} - \SI{2.5}{\milli\metre}, LLR still lags behind VLBI, however its long time span and the possibility of dynamic realisation can be beneficial for some applications. Additionally, the results from LLR can be used to validate the results from VLBI.

\begin{table}[!ht]
\caption{Mean values of the accuracy of $\Delta$UT values obtained from different subsets (\enquote*{Mean 3$\sigma$} in table), and the mean of absolute changes to $\Delta$UT values (i.e. differences of estimated values to the values from the IERS 14 C04 series (\enquote*{Mean diff.} in table)) for the results from the beginning of any dataset until 2000.0 and the results after 2000.0.}\label{tab:dut_subsets}
\centering
\begin{tabular}{cP{1cm}P{1cm}P{1cm}P{1cm}}
\noalign{\smallskip}\hline\noalign{\smallskip}
& \multicolumn{2}{P{2cm}}{Before 2000.0 [\si{\micro\second}]} & \multicolumn{2}{P{2cm}}{After 2000.0 [\si{\micro\second}]} \\
\noalign{\smallskip}\cline{2-5}\noalign{\smallskip}
Subset & Mean 3$\sigma$ & Mean diff. & Mean 3$\sigma$ & Mean diff. \\
\noalign{\smallskip}\hline\noalign{\smallskip}
Apollo05 & - & - & 14.96 & 116.31 \\ \noalign{\smallskip}\hline \noalign{\smallskip}
Apollo10 & - & - & 15.92 & 98.99 \\
\noalign{\smallskip} \hline \noalign{\bigskip}
OCA05 & 96.31 & 204.35 & 30.71 & 69.07 \\ \noalign{\smallskip}\hline \noalign{\smallskip}
OCA10 & 71.06 & 155.32 & 20.27 & 50.04 \\ \noalign{\smallskip}\hline \noalign{\smallskip}
OCA15 & 53.56 & 128.89 & 16.53 & 39.95 \\ \noalign{\smallskip} \hline \noalign{\bigskip}
All05 & 115.52 & 253.96 & 24.79 & 82.88 \\ \noalign{\smallskip}\hline \noalign{\smallskip}
All10 & 74.13 & 169.21 & 18.63 & 56.91 \\ \noalign{\smallskip}\hline \noalign{\smallskip}
All15 & 63.56 & 154.12 & 15.89 & 44.27 \\
\noalign{\smallskip}\hline
\end{tabular}
\end{table}

\subsection{Correlations}\label{subsec:corr_dut}
The correlations of different parameters which are estimated in LUNAR with the ERP of the different estimated nights (henceforth, only \enquote*{correlations}, in this section) were briefly discussed by \citet{bisk2015}. In this study, as mentioned before, the velocities of the LLR observatories were kept fixed to the ITRF2014 solution values, as they have shown high correlations with the estimated ERP in previous studies.

One of the benefits of estimating the ERP from only one observatory (compared to estimating from all observatories) is that it leads to a reduction in the correlations estimated within LUNAR, as also visible by Table \ref{tab:corr_dut}, where we show the value of the maximum correlation of $\Delta$UT for any night with any non-ERP parameter, obtained from each subset.


\begin{table}[!htp]
\caption{The maximum correlation of $\Delta$UT of any night with any non-ERP estimated parameter, obtained from each subset.}\label{tab:corr_dut}
\centering
\begin{tabular}{ccccc}
\hline\noalign{\smallskip}
Subset name prefix &\multicolumn{3}{c}{Subset name suffix} \\
 &05 &10 &15 \\
\noalign{\smallskip}\hline\noalign{\smallskip}
Apollo  & \SI{50}{\percent} & \SI{30}{\percent} & - \\
\noalign{\smallskip}\hline\noalign{\smallskip}
OCA & \SI{40}{\percent} & \SI{20}{\percent} & \SI{20}{\percent} \\
\noalign{\smallskip}\hline\noalign{\smallskip}
All & \SI{80}{\percent} & \SI{70}{\percent} & \SI{30}{\percent} \\
\noalign{\smallskip}\hline
\end{tabular}
\end{table}

Overall, most parameters show no or very low correlation with the $\Delta$UT values of any night. For the subsets from all observatories, the highest correlation, from all three subsets, of \SI{80}{\percent}, \SI{70}{\percent}, and \SI{30}{\percent} is with the y-coordinate of WLRS. The parameters, from all three \enquote*{All} subsets, which show more than a \SI{30}{\percent} correlation with the $\Delta$UT values of any night are: the coordinates and/or biases of APOLLO, MLRS, LURE, OCA, and WLRS. Additionally, the x-coordinate of the L1 reflector shows a \SI{30}{\percent} correlation only for the subset All05. For the subsets from OCA, the parameters (from all three OCA subsets) which show more than a \SI{20}{\percent} correlation with the $\Delta$UT values of any night are only the coordinates and biases of OCA. For the subsets from APOLLO, the parameters (from both APOLLO subsets) which show more than a \SI{30}{\percent} correlation with the $\Delta$UT values of any night are the coordinates and biases of APOLLO. Additionally, for the subset Apollo05, the x-coordinate of OCA and the x-coordinate of the L1 reflector shows a \SI{30}{\percent} correlation.

Overall, the correlation between $\Delta$UT values and the positions of LLR observatories is as expected. Additionally, the correlations with parameters such as the coordinates of lunar reflector disappear (or become smaller, and therefore irrelevant) when implementing the stricter selection criteria (10 or 15 \acp{NP} per night)

\subsection{Effect of NTL on \texorpdfstring{$\Delta$}{}UT}\label{subsec:ntsl_dut}
\begin{figure}[!ht]
    \centering
    \includegraphics[width=0.49\textwidth]{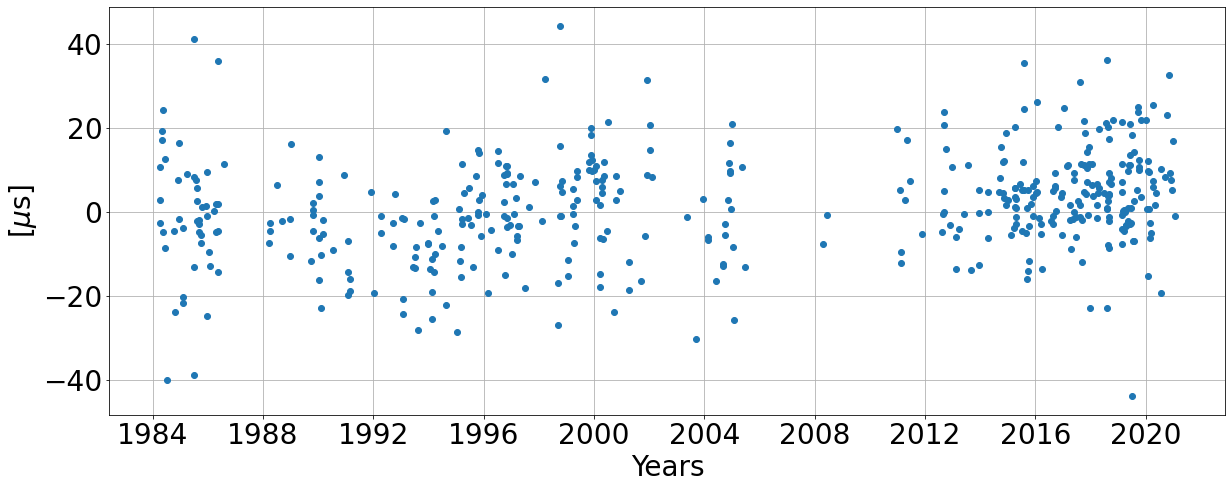}
    \caption{Estimated values of $\Delta$UT for the NTSL solution from IMLS subtracted from the estimated values of the standard solution for the subset All15.}\label{fig:ntsl_all15all_dut}
\end{figure}

As there are correlations between the station coordinates and $\Delta$UT, and as the addition of \ac{NTL} leads to small benefits in our previous study, we decided to test the effect of including \ac{NTL} in the ERP determination from LUNAR. Here, we added \ac{NTL} only from IMLS, as it performed the best in LLR analysis \citep{singh_etal_2021}. We represent the combination of its three components (NTAL, NTOL, and HYDL) as NTSL. The effect of \ac{NTL}, as expected, is of similar - small - magnitude on the estimated $\Delta$UT values from all subsets for which the calculations were performed. In Fig. \ref{fig:ntsl_all15all_dut}, we show the effect when adding NTSL to the standard solution for the estimation of $\Delta$UT from the subset All15. Additionally, in Table \ref{tab:ntsl_dut}, we show the mean values of the accuracy obtained for the estimation of $\Delta$UT from the subsets All10 and All15 for the standard solution and the \ac{NTL} solutions from IMLS. For the both the subsets, the NTSL solution improves the accuracy by \SI{0.9}{\percent}.

\begin{table}
\caption{Mean values of the accuracy of $\Delta$UT for the standard and the \ac{NTL} solutions for the subsets All10 and All15.}\label{tab:ntsl_dut}
\centering
\begin{tabular}{cccc}
\hline\noalign{\smallskip}
&Loading &{All10 [\si{\micro\second}]} &{All15 [\si{\micro\second}]} \\
\noalign{\smallskip}\hline\noalign{\smallskip}
Std &  & 42.31 &35.80 \\
\noalign{\smallskip}\hline\noalign{\smallskip}

\multirow{4}[6]{*}{IMLS} &NTAL &42.15 &35.68  \\
\noalign{\smallskip}\cline{2-4}\noalign{\smallskip}
&NTOL &42.23 &35.74 \\
\noalign{\smallskip}\cline{2-4}\noalign{\smallskip}
&HYDL &42.11 &35.64  \\
\noalign{\smallskip}\cline{2-4}\noalign{\smallskip}
&NTSL &41.92 &35.47 \\
\noalign{\smallskip}\hline
\noalign{\smallskip}

\end{tabular}
\end{table}



\begin{figure}[!ht]
    \centering
    \includegraphics[width=0.49\textwidth]{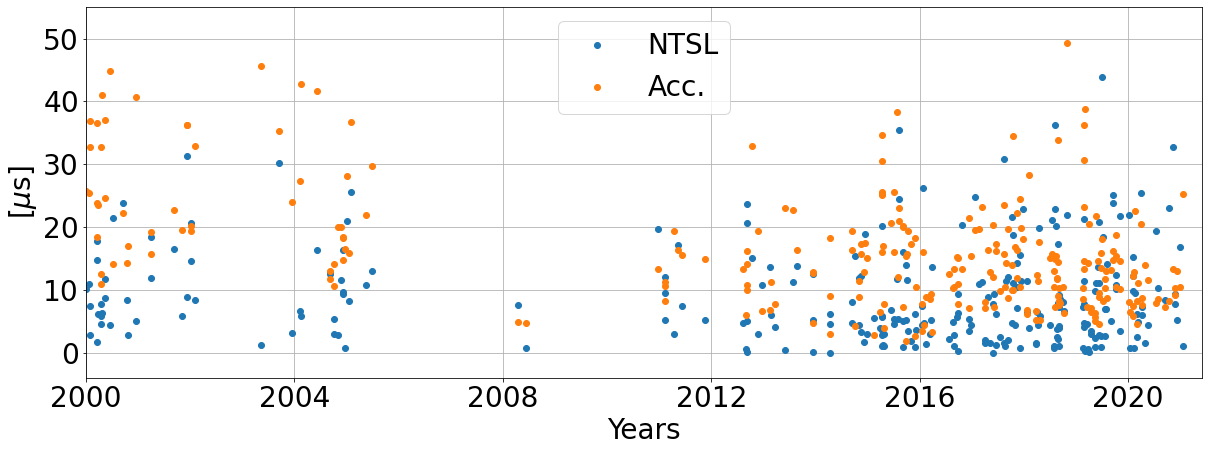}
    \caption{Absolute values of the effect when adding of NTSL on $\Delta$UT values (in blue) and the accurcaies of the estimated $\Delta$UT values (in orange) for the subset All15, after 2000.0.}\label{fig:ntsl_all15all_acc_dut}
\end{figure}

As the effect of adding \ac{NTL} is similar for the whole time span, its effect on the results of the estimation of $\Delta$UT does not change significantly when considering the entire time span of any subset, or only the time span which includes NPs with higher accuracy. Additionally, the inclusion of NTSL does not change the correlations, compared to the standard solution, at all. In Fig. \ref{fig:ntsl_all15all_acc_dut}, we show a comparison of the effect of NTSL on on $\Delta$UT values (absolute values) with the accuracies of the estimation of $\Delta$UT values for the results after 2000.0 (for the subset All15), as the accuracies in this time span are significantly better than before 2000.0. It can be seen that both, the absolute effect of NTSL on $\Delta$UT and the accuracies of the $\Delta$UT values, are in a similar range. For the subset All15, shown in Fig. \ref{fig:ntsl_all15all_acc_dut}, the mean of the effect of NTSL (absolute values) is \SI{9.13}{\micro\second}, and the mean of the accuracy of $\Delta$UT is \SI{15.89}{\micro\second} (see Table \ref{tab:dut_subsets}).  Even though its effect is not significantly larger than accuracies obtained of $\Delta$UT values, we assess that there is a small improvement in the overall accuracy of the estimated $\Delta$UT values when including the NTSL. The improvement of the results due to the addition of IMLS NTSL is in sync with the findings from our previous study \citep{singh_etal_2021}.

\section{Estimation of terrestrial pole offsets}\label{sec:xpyp}
The PMC were estimated (for the nights of all subsets, Table \ref{tab:subsets_nights}), using a-priori values from the combination of KEOF COMB2019 and IERS 14 C04 series, as mentioned previously. In the subsections that follow, we discuss the results of the estimated values of $x_p$ and $y_p$ and their accuracy, their correlations with each other and with other parameters, and the effect of \ac{NTL} on them.

For the estimation of $x_p$ and $y_p$, the velocities of the LLR observatories were fixed to the ITRF 2014 solution values, as it was also done for $\Delta$UT estimation (see section \ref{sec:dUT}), and the values of $\Delta$UT were fixed to their a-priori values.



\subsection{Estimated values}\label{subsec:res_xpyp}
\begin{figure*}[!ht]
    \begin{subfigure}{.48\textwidth}
    \centering
    \includegraphics[width=1\textwidth]{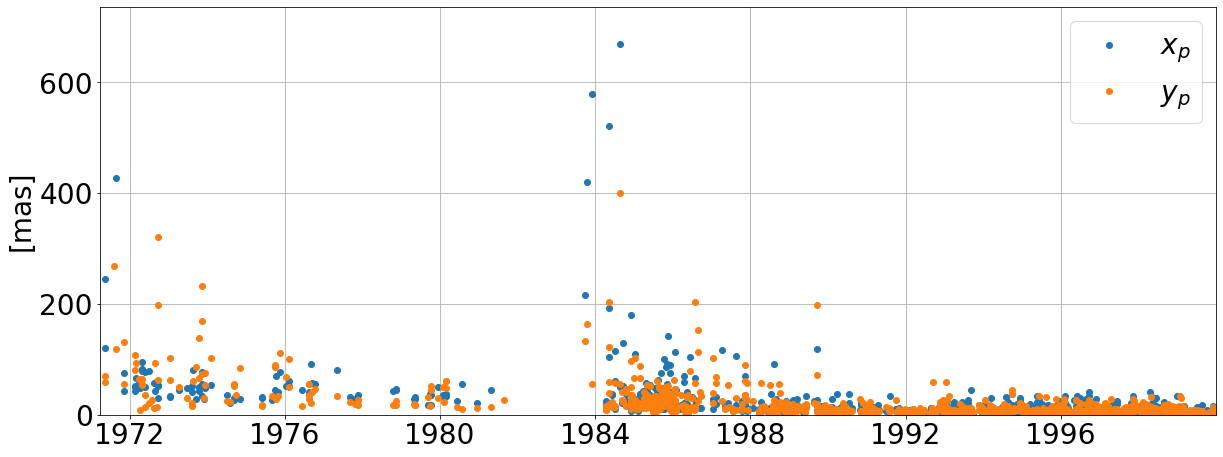}
    \caption{All05}
    \end{subfigure}%
    \qquad
    \begin{subfigure}{.48\textwidth}
    \centering
    \includegraphics[width=1\textwidth]{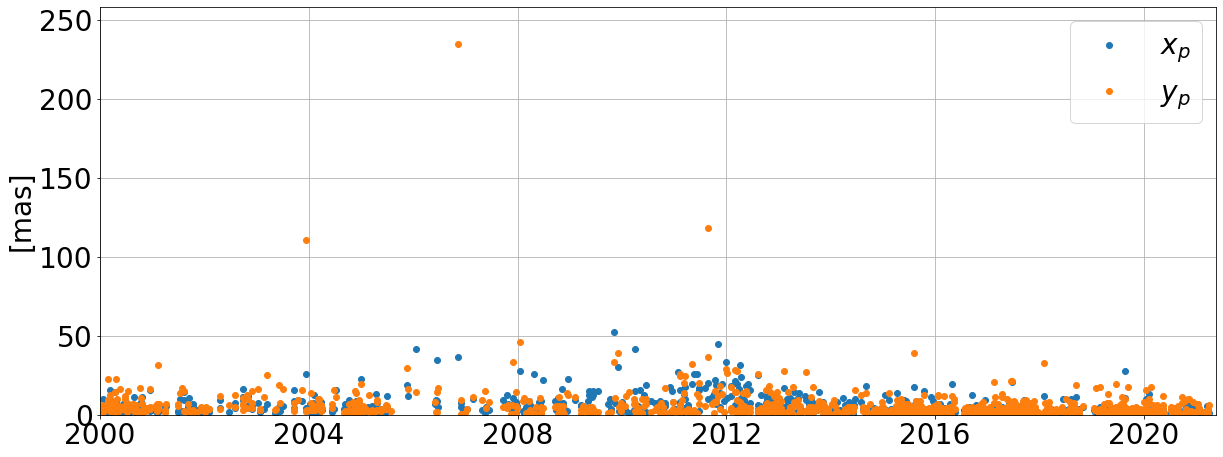}
    \caption{All05}
    \end{subfigure}
    
    \begin{subfigure}{.48\textwidth}
    \centering
    \includegraphics[width=1\textwidth]{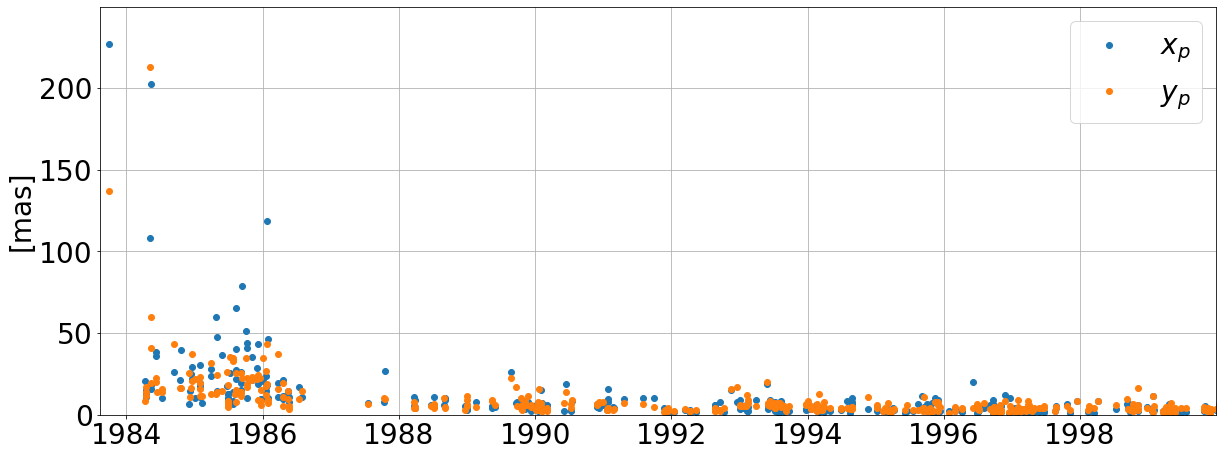}
    \caption{All10}
    \end{subfigure}%
    \qquad
    \begin{subfigure}{.48\textwidth}
    \centering
    \includegraphics[width=1\textwidth]{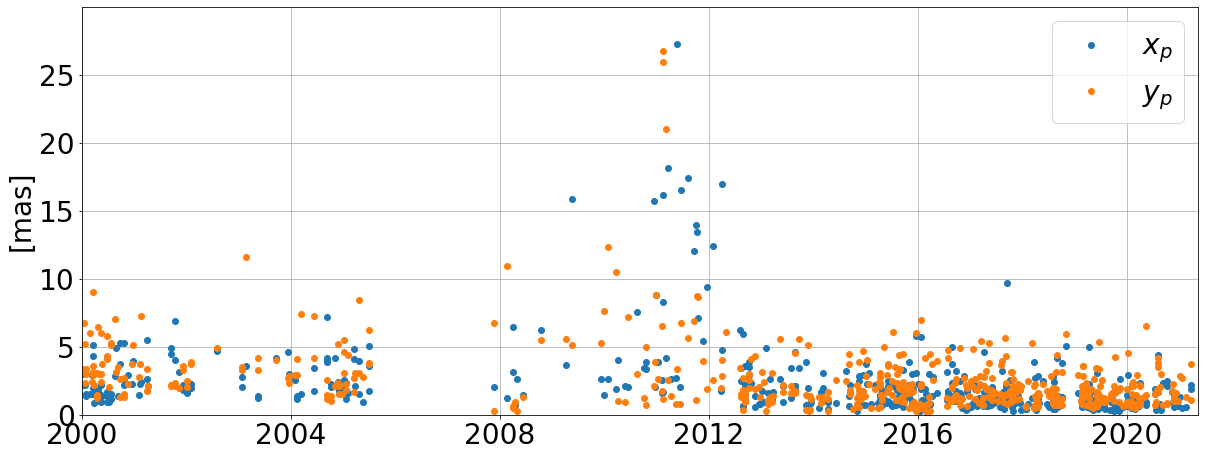}
    \caption{All10}
    \end{subfigure}
    
    \begin{subfigure}{.48\textwidth}
    \centering
    \includegraphics[width=1\textwidth]{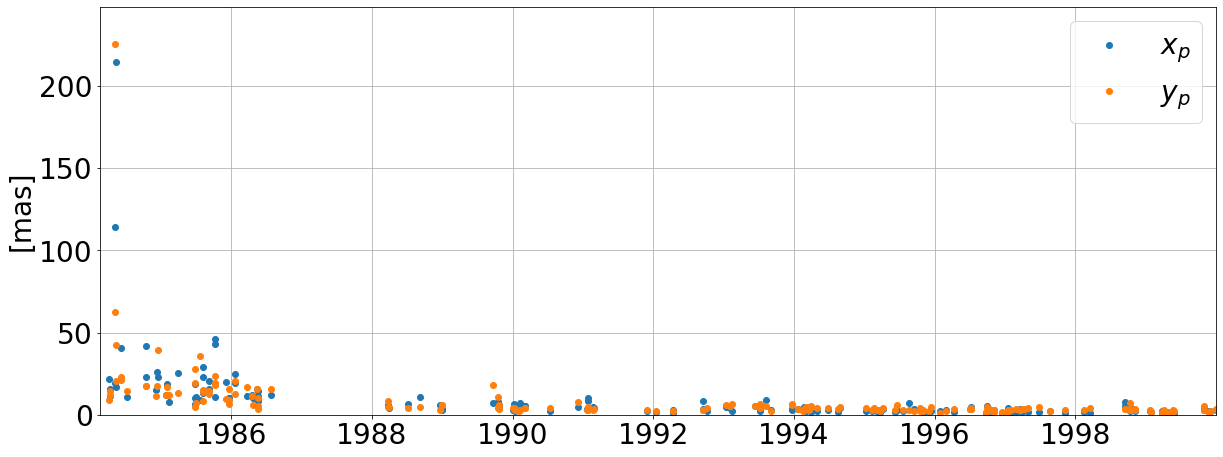}
    \caption{All15}
    \end{subfigure}%
    \qquad
    \begin{subfigure}{.48\textwidth}
    \centering
    \includegraphics[width=1\textwidth]{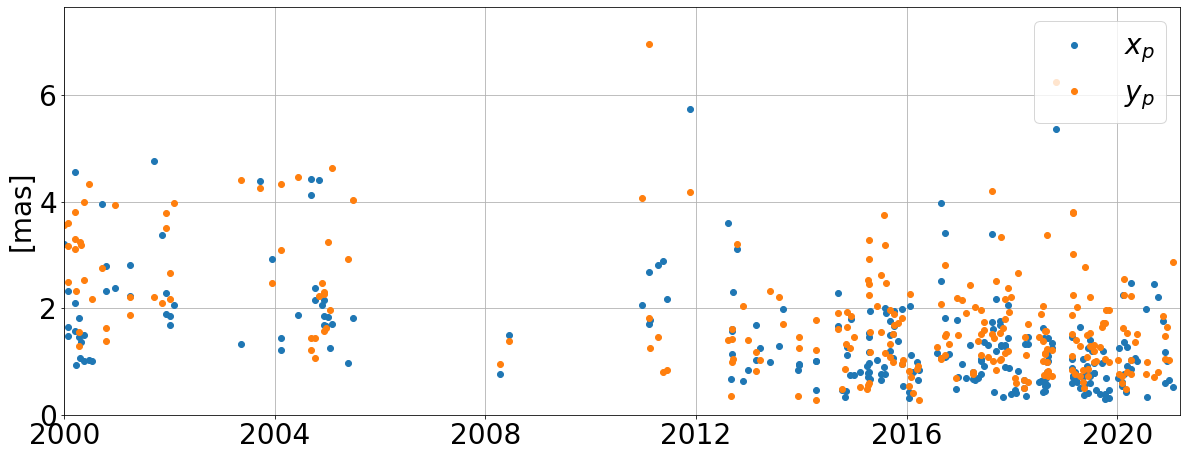}
    \caption{All15}
    \end{subfigure}
    \caption{Accuracy of the estimated values of PMC for the subset All05, All10, and All15: in (a), (c), and (e) from the beginning of the dataset until 2000.0, and in (b), (d), and (f) from 2000.0 until the end of the dataset. Note the differences in the range of the axes for each sub-figure.}
    \label{fig:all_pol}
\end{figure*}

The estimation of $x_p$ and $y_p$ (following the GMM) gives the possibility of estimating $x_p$ and $y_p$ individually, or together. In the previous results of $x_p$ and $y_p$ estimation from LUNAR \citep{bisk2015}, the estimation was done individually, as the correlation between the $x_p$ and $y_p$ values (of the same night) is high, and the adjusted values were deemed to be more realistic when the adjustment was performed individually. With the expansion of the LLR dataset over the past years (providing a longer time series), more NPs recorded per night, and improved coverage of the lunar orbit mainly because of the benefits using IR laser pulses from OCA, the estimation of $x_p$ and $y_p$ together from has benefited. In this study, we estimate the PMC together and separately, for all subsets mentioned in Table \ref{tab:subsets_nights}, and compare the results. In Table \ref{tab:pol_subsets}, we give the mean values of the accuracy of $x_p$ and $y_p$ achieved, and the differences of the adjusted values to the IERS 14 C04 series. Additionally, as we did for $\Delta$UT values, we split the results of PMC into two time spans, setting a break at 2000.0, for each subset, due to the stark improvement in the involved NPs and the results obtained over the LLR time span.

Fig. \ref{fig:all_pol} shows the accuracy of the PMC values for the subsets All05, All10, and All15, also split into two time spans, with a break at 2000.0. Both, the figure and Table \ref{tab:pol_subsets}, give manifold results, discussed below. 

As the x-axis for the polar motion is defined by the Greenwich meridian, and the the y-axis is defined by the line joining the \SI{90}{\degree} meridian (see Fig. 2 of \cite{Combrinck09}), OCA is (and other European observatories would be) more sensitive to the estimation of $x_p$ compared to $y_p$, whereas APOLLO is (and other American observatories would be) more sensitive to the estimation of $y_p$ compared to $x_p$. This can also be seen by Table \ref{tab:pol_subsets}, where the mean accuracy of the estimated $x_p$ values is better than the mean of the estimated $y_p$ values, for the results after 2000.0 (for the \enquote*{OCA} and \enquote*{All} subsets). This is, however, not visible for the results before 2000.0, probably because of the combination of the worse accuracy of the NPs involved along with the high sensitivity of OCA to the x-axis. The results of $x_p$ show a higher improvement between the split subsets, compared to $y_p$ (for example, mean values of $x_p$ improves by \SI{80.79}{\percent} from \SI{9.42}{\mas} to \SI{1.81}{\mas} for the subset OCA10, whereas the mean value of $y_p$ improves by \SI{70.89}{\percent} from \SI{8.69}{\mas} to \SI{2.53}{\mas}). As described in section \ref{sec:llr_data}, the LLR data is not evenly distributed with OCA contributing to over \SI{60}{\percent} of the NPs, leading to better estimation of $x_p$ from the subsets which consider NPs from all observatories as well, as most nights in the three subsets from all observatories are from OCA.

Other than to avoid high correlation between the estimated PMC of the same night, the individual estimation of $x_p$ and $y_p$ also helps see whether the estimation of both $x_p$ and $y_p$ is stable or not. As seen in Table \ref{tab:pol_subsets}, the mean values of the accuracies of $x_p$ and $y_p$ change the most for the subsets made from selection of 5 NPs per night, and change the least for selection of 15 NPs per night, indicating that the calculations with selection of only 5 NPs per night are not as stable as those with a higher number of NPs selected per night. Additionally, it can be seen that the results for the subsets from APOLLO change drastically when estimating $x_p$ and $y_p$ individually, compared to their estimation together, even though the NPs from APOLLO are the most accurate, amongst all NPs. This indicates that the estimation of the PMC is more stable when the number of nights for which the estimation is performed is higher.

Table \ref{tab:pol_subsets} also gives the differences obtained between the PMC values obtained from LUNAR and the IERS 14 C04 series. These differences follow the same trend as the accuracies, i.e. the results with the best accuracies have the least differences to the C04 series. These differences, as was also the case for differences of $\Delta$UT from the C04 series, are presumed to be due to the different space geodetic techniques involved in formation and combination of the final product. Additionally, they could also be caused by some systematic errors in our calculation. 

For the resolution of PMC, \SI{1}{\mas} corresponds to \SI{3}{\centi\metre} on the Earth's surface, implying that the current and best possible resolution of the PMC from \ac{LLR} (subset All15, after 2000.0) is at \SI{3.5}{\centi\metre} for $x_p$ and at \SI{4.6}{\centi\metre} for $y_p$ on Earth's surface. Compared to the resolution obtained from other space geodetic techniques, such as GNSS, the results of PMC estimation from LLR lag far behind.

\begin{table*}[!ht]
\caption{Mean values of the accuracies of the $x_p$ and $y_p$ values obtained from different subsets (\enquote*{3$\sigma$} in table), and the mean of absolute changes to $x_p$ and $y_p$ values (i.e. differences of estimated values to the values from the IERS 14 C04 series (\enquote*{Diff.} in table)) for the results from the beginning of the dataset until 2000.0 in (a) and from 2000.0 until the end of the dataset in (b).}\label{tab:pol_subsets}
\begin{subtable}{1.0\textwidth}
\centering
\caption{Before 2000.0}
\begin{tabular}{P{1.3cm}P{1cm}P{1cm}P{1cm}P{1cm}P{1cm}P{1cm}P{1cm}P{1cm}P{1cm}}
\hline\noalign{\smallskip}
Subset &\multicolumn{4}{c}{$x_p$ and $y_p$ [\si{\mas}]} &\multicolumn{2}{c}{Only $x_p$ [\si{\mas}]}   &\multicolumn{2}{c}{Only $y_p$ [\si{\mas}]} \\
&\multicolumn{2}{c}{3$\sigma$} &\multicolumn{2}{c}{Diff.} &3$\sigma$ &Diff. &3$\sigma$ &Diff. \\
&$x_p$ & $y_p$ &$x_p$ & $y_p$ & $x_p$ & $x_p$ &$y_p$ &$y_p$ \\
\noalign{\smallskip}\hline\noalign{\smallskip}
Apollo05 & - & - & - & - & - & - & - & - \\
\noalign{\smallskip}\hline\noalign{\smallskip}
Apollo10 & - & - & - & - & - & - & - & - \\
\noalign{\smallskip}\hline\noalign{\bigskip}
OCA05 & 13.79 & 13.68 & 3.75 & 4.07 & 9.91 & 3.26 & 9.63 & 3.29 \\
\noalign{\smallskip}\hline\noalign{\smallskip}
OCA10 & 9.42 & 8.69 & 2.62 & 2.41 & 7.89 & 2.73 & 7.19 & 2.52 \\
\noalign{\smallskip}\hline\noalign{\smallskip}
OCA15 & 7.39 & 5.95 & 2.55 & 2.16 & 6.75 & 2.54 & 5.29 & 2.16 \\
\noalign{\smallskip}\hline\noalign{\bigskip}
All05 & 19.92 & 18.56 & 10.33 & 8.85 & 13.17 & 6.22 & 12.60 & 5.81 \\
\noalign{\smallskip}\hline\noalign{\smallskip}
All10 & 9.66 & 8.73 & 2.93 & 2.65 & 8.10 & 3.06 & 6.96 & 2.48 \\
\noalign{\smallskip}\hline\noalign{\smallskip}
All15 & 8.57 & 8.06 & 2.76 & 2.82 & 7.20 & 2.23 & 6.51 & 2.29 \\
\noalign{\smallskip}\hline\noalign{\smallskip}
\end{tabular}

\end{subtable}

\begin{subtable}{1.0\textwidth}
\centering
\caption{After 2000.0}
\begin{tabular}{P{1.3cm}P{1cm}P{1cm}P{1cm}P{1cm}P{1cm}P{1cm}P{1cm}P{1cm}P{1cm}}
\hline\noalign{\smallskip}
Subset &\multicolumn{4}{c}{$x_p$ and $y_p$ [\si{\mas}]} &\multicolumn{2}{c}{Only $x_p$ [\si{\mas}]}   &\multicolumn{2}{c}{Only $y_p$ [\si{\mas}]} \\
&\multicolumn{2}{c}{3$\sigma$} &\multicolumn{2}{c}{Diff.} &3$\sigma$ &Diff. &3$\sigma$ &Diff. \\
&$x_p$ & $y_p$ &$x_p$ & $y_p$ & $x_p$ & $x_p$ &$y_p$ &$y_p$ \\
\noalign{\smallskip}\hline\noalign{\smallskip}
Apollo05 & 8.41 & 6.88 & 9.13 & 5.89 & 2.22 & 2.64 & 1.48 & 1.63 \\
\noalign{\smallskip}\hline\noalign{\smallskip}
Apollo10 & 5.89 & 4.55 & 3.85 & 2.61 & 2.19 & 2.40 & 1.31 & 1.37 \\
\noalign{\smallskip}\hline\noalign{\bigskip}
OCA05 & 3.42 & 5.08 & 1.44 & 1.73 & 1.97 & 0.96 & 2.97 & 1.11 \\
\noalign{\smallskip}\hline\noalign{\smallskip}
OCA10 & 1.81 & 2.53 & 0.78 & 0.68 & 1.38 & 0.79 & 1.97 & 0.78 \\
\noalign{\smallskip}\hline\noalign{\smallskip}
OCA15 & 1.39 & 1.85 & 0.74 & 0.56 & 1.20 & 0.76 & 1.62 & 0.70 \\
\noalign{\smallskip}\hline\noalign{\bigskip}
All05 & 4.49 & 5.37 & 3.25 & 2.87 & 1.99 & 1.37 & 2.59 & 1.28 \\
\noalign{\smallskip}\hline\noalign{\smallskip}
All10 & 2.18 & 2.54 & 1.19 & 0.89 & 1.38 & 0.99 & 1.74 & 0.83 \\
\noalign{\smallskip}\hline\noalign{\smallskip}
All15 & 1.38 & 1.77 & 0.81 & 0.59 & 1.16 & 0.83 & 1.52 & 0.73 \\
\noalign{\smallskip}\hline\noalign{\smallskip}
\end{tabular}

\end{subtable}

\end{table*}

\subsection{Correlations}\label{subsec:corr_pol}

\begin{figure}[!ht]
    \begin{subfigure}{.48\textwidth}
    \centering
    \includegraphics[width=1\textwidth]{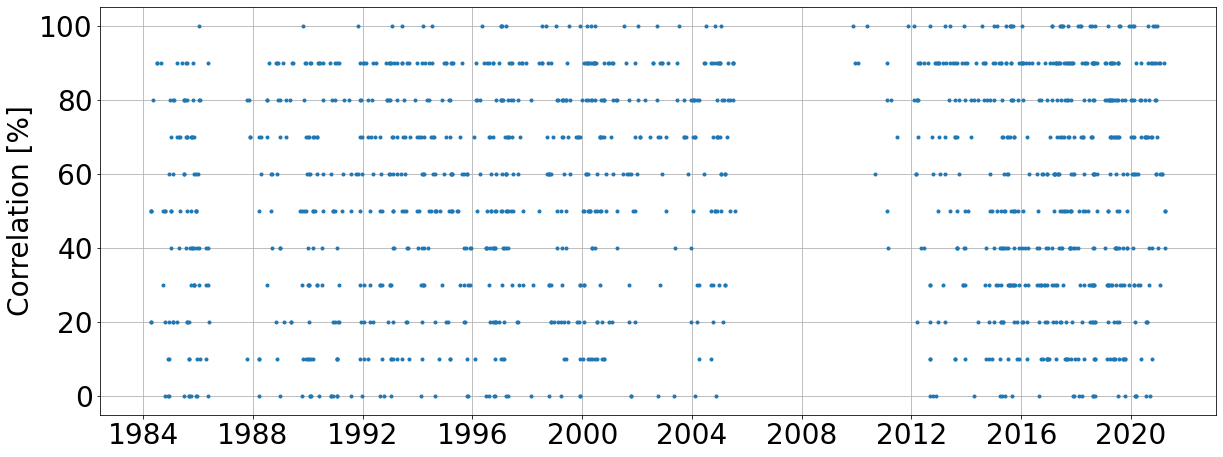}
    \caption{OCA05}
    \end{subfigure}%
    \\
    \\
    \begin{subfigure}{.48\textwidth}
    \centering
    \includegraphics[width=1\textwidth]{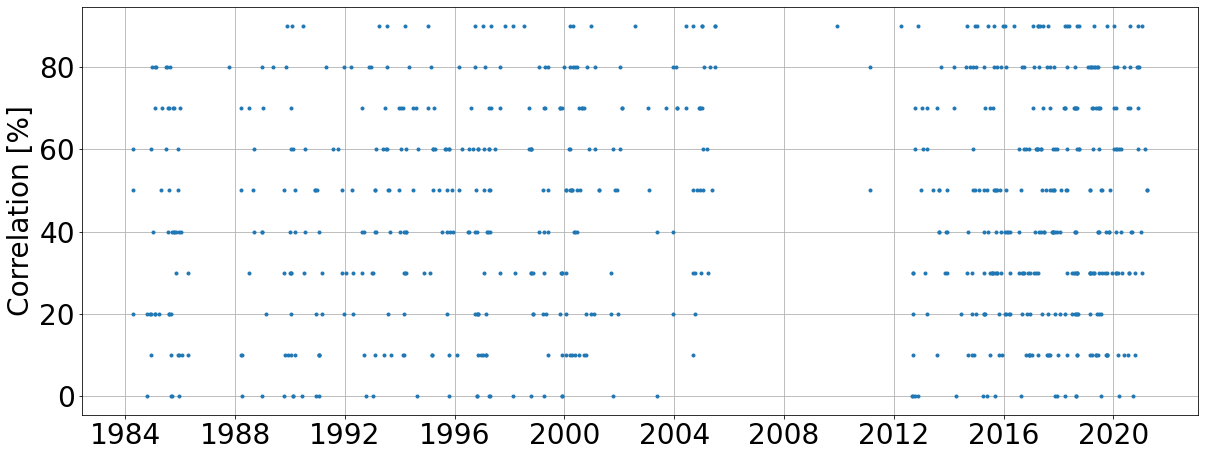}
    \caption{OCA10}
    \end{subfigure}
    \\
    \\
    \begin{subfigure}{.48\textwidth}
    \centering
    \includegraphics[width=1\textwidth]{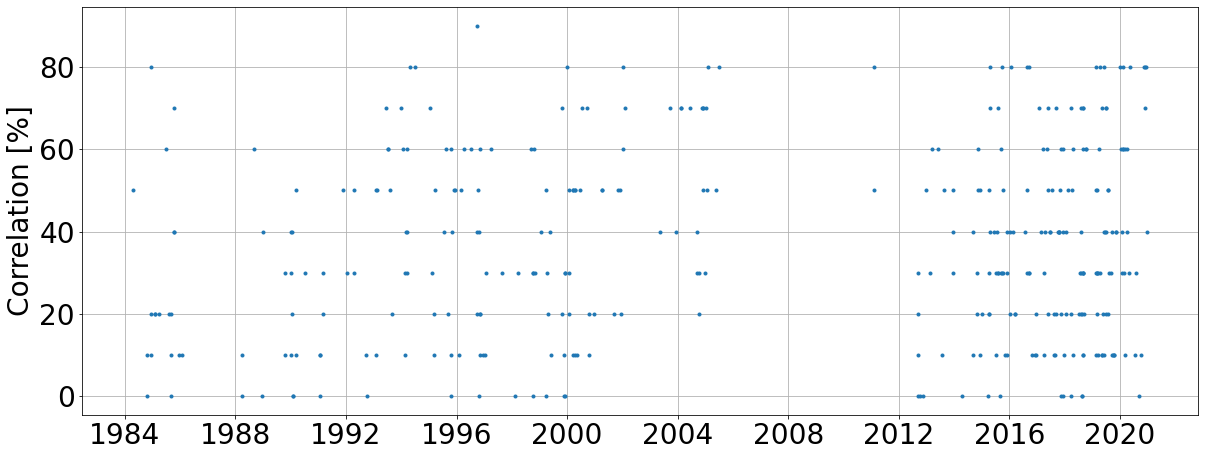}
    \caption{OCA15}
    \end{subfigure}%
    
    \caption{Correlation of $x_p$ and $y_p$ estimated in the same night with each other, from the subsets OCA05, OCA10, and OCA15.}
    \label{fig:corr_pol_xpyp}
\end{figure}

As mentioned before, when estimating the PMC together, the values of $x_p$ and $y_p$ in the same night are correlated to each other. These correlations can be as high as \SI{100}{\percent}, indicating advantage of estimating the PMC separately. In Fig. \ref{fig:corr_pol_xpyp}, we give the correlation of the $x_p$ and $y_p$ estimated in the same night with each other, from the subsets OCA05, OCA10, and OCA15. The trend followed by the other subsets is the same, i.e., the correlations of the subsets chosen with 5 NPs per night are the highest, with the PMC of many nights showing a \SI{100}{\percent} correlation with each other. With the selection of 15 NPs per night, none of the nights (for OCA15) show a correlation of \SI{100}{\percent} with each other, further proving the benefit of the strict selection criteria of subsets for ERP estimation from LLR. However, even with 15 NPs per night, the correlations of the $x_p$ and $y_p$ estimated in the same night with each other are very high.

For the correlation of the estimated $x_p$ and $y_p$ with the non-ERP parameters of LUNAR, the trend followed is the same as that for the estimation of $\Delta$UT values, i.e., the correlation of the estimation from \enquote*{All} subsets is higher than the correlation of the estimation of ERP from nights selected from any one observatory. In Table \ref{tab:corr_pol}, we show the value of the maximum correlations of $x_p$ and $y_p$ of any night with any non-ERP parameter for each subset.


As it was the case with correlations of $\Delta$UT with non-ERP parameters, most parameters show either no or very low correlation with the estimated $x_p$ and $y_p$. The parameters which show the maximum correlations, whether the PMC are estimated together or separately, stay the same. The highest correlations from the subset All05 and All10 are with the position of MAUI and MLRS. For the subset All15, the highest correlations are with the position of WLRS. Overall, the parameters which show a \SI{40}{\percent} correlation or more with either $x_p$ and $y_p$ are the station coordinates and/or biases of the LLR observatories. For OCA05 and OCA10 subsets, the highest correlation reach up to \SI{50}{\percent}, where, \SI{30}{\percent} or higher correlation is achieved only by the coordinates and/or biases of OCA. For Apollo05, the maximum correlations reach up to \SI{60}{\percent}, where the parameters which show a \SI{30}{\percent} or higher correlation are only the coordinates and/or biases of APOLLO. For Apollo10, only the coordinates and/or biases of APOLLO show a \SI{20}{\percent} or more correlation, reaching a maximum of \SI{40}{\percent}.

The parameters which show a correlation with the PMC are the same, whether the estimation is done separately or together. The percentage by which they are correlated also stays very similar, as can also be seen in Table \ref{tab:corr_pol} (specially for the subsets with the strict selection criteria of 15 NPs per night). The difference in correlations between these two cases is that more nights show a correlation with those parameters when the estimation of PMC is done separately. Some effect is also absorbed by the correlation between the $x_p$ and the $y_p$ of the same night.

\begin{table*}[!htp]
\caption{The maximum correlations of $x_p$ and $y_p$ of any night with any non-ERP estimated parameter, obtained from each subset.}
\centering
\begin{subtable}{.5\textwidth}
\centering
\caption{$x_p$, estimation of PMC together}
\begin{tabular}{ccccc}
\hline\noalign{\smallskip}
Subset name prefix &\multicolumn{3}{c}{Subset name suffix} \\
 &05 &10 &15 \\
\noalign{\smallskip}\hline\noalign{\smallskip}
Apollo  & \SI{60}{\percent} & \SI{40}{\percent} & - \\
\noalign{\smallskip}\hline\noalign{\smallskip}
OCA & \SI{50}{\percent} & \SI{30}{\percent} & \SI{30}{\percent} \\
\noalign{\smallskip}\hline\noalign{\smallskip}
All & \SI{100}{\percent} & \SI{100}{\percent} & \SI{40}{\percent} \\
\noalign{\smallskip}\hline\noalign{\smallskip}\noalign{\smallskip}
\end{tabular}

\end{subtable}
\begin{subtable}{.5\textwidth}
\centering
\caption{$y_p$, estimation of PMC together}
\begin{tabular}{ccccc}
\hline\noalign{\smallskip}
Subset name prefix &\multicolumn{3}{c}{Subset name suffix} \\
 &05 &10 &15 \\
\noalign{\smallskip}\hline\noalign{\smallskip}
Apollo  & \SI{60}{\percent} & \SI{30}{\percent} & - \\
\noalign{\smallskip}\hline\noalign{\smallskip}
OCA & \SI{30}{\percent} & \SI{20}{\percent} & \SI{10}{\percent} \\
\noalign{\smallskip}\hline\noalign{\smallskip}
All & \SI{90}{\percent} & \SI{90}{\percent} & \SI{30}{\percent} \\
\noalign{\smallskip}\hline\noalign{\smallskip}\noalign{\smallskip}
\end{tabular}

\end{subtable}

\begin{subtable}{.5\textwidth}
\centering
\caption{$x_p$, estimation of only $x_p$}
\begin{tabular}{ccccc}
\hline\noalign{\smallskip}
Subset name prefix &\multicolumn{3}{c}{Subset name suffix} \\
 &05 &10 &15 \\
\noalign{\smallskip}\hline\noalign{\smallskip}
Apollo  & \SI{50}{\percent} & \SI{30}{\percent} & - \\
\noalign{\smallskip}\hline\noalign{\smallskip}
OCA & \SI{50}{\percent} & \SI{30}{\percent} & \SI{30}{\percent} \\
\noalign{\smallskip}\hline\noalign{\smallskip}
All & \SI{80}{\percent} & \SI{80}{\percent} & \SI{40}{\percent} \\
\noalign{\smallskip}\hline\noalign{\smallskip}
\end{tabular}

\end{subtable}
\begin{subtable}{.5\textwidth}
\centering
\caption{$y_p$, estimation of only $y_p$}
\begin{tabular}{ccccc}
\hline\noalign{\smallskip}
Subset name prefix &\multicolumn{3}{c}{Subset name suffix} \\
 &05 &10 &15 \\
\noalign{\smallskip}\hline\noalign{\smallskip}
Apollo  & \SI{60}{\percent} & \SI{30}{\percent} & - \\
\noalign{\smallskip}\hline\noalign{\smallskip}
OCA & \SI{40}{\percent} & \SI{20}{\percent} & \SI{10}{\percent} \\
\noalign{\smallskip}\hline\noalign{\smallskip}
All & \SI{70}{\percent} & \SI{60}{\percent} & \SI{40}{\percent} \\
\noalign{\smallskip}\hline\noalign{\smallskip}
\end{tabular}

\end{subtable}
\label{tab:corr_pol}
\end{table*}

\subsection{PMC signal analysis}\label{subsec:signals_pol}
The polar motion has three major components: Chandler wobble, annual oscillation, and a drift along the \SI{80}{\degree} West meridian. When applying a Fourier transformation of the PMC from the IERS 14 C04 series (not shown), signals with an annual period and a Chandler period are visible. As the LLR NPs are temporally unevenly distributed, a Fourier transformation of the estimated PMC values is not possible. To perform a spectral analysis of a non-uniformly distributed data, we used the Lomb-Scargle (LS) periodogram. However, to obtain a very clear distribution, a high sampling rate and uniformity of data samples is beneficial \citep{ls_ana}, which is not given in LLR dataset. Due to these reasons, the LS power (not shown) at the annual and Chandler frequency component did not show similar high peaks as the Fourier transformation of the PMC from the IERS 14 C04 series for all subsets. In the difference of the powers obtained from the LS periodograms of the a-priori PMC values and the estimated PMC values from LUNAR, no change of the signals was visible when estimating PMC (whether together or separately). 

\subsection{Effect of NTL on estimated values}\label{subsec:ntsl_xpyp}
\begin{figure}[!ht]
    \centering
    \includegraphics[width=0.49\textwidth]{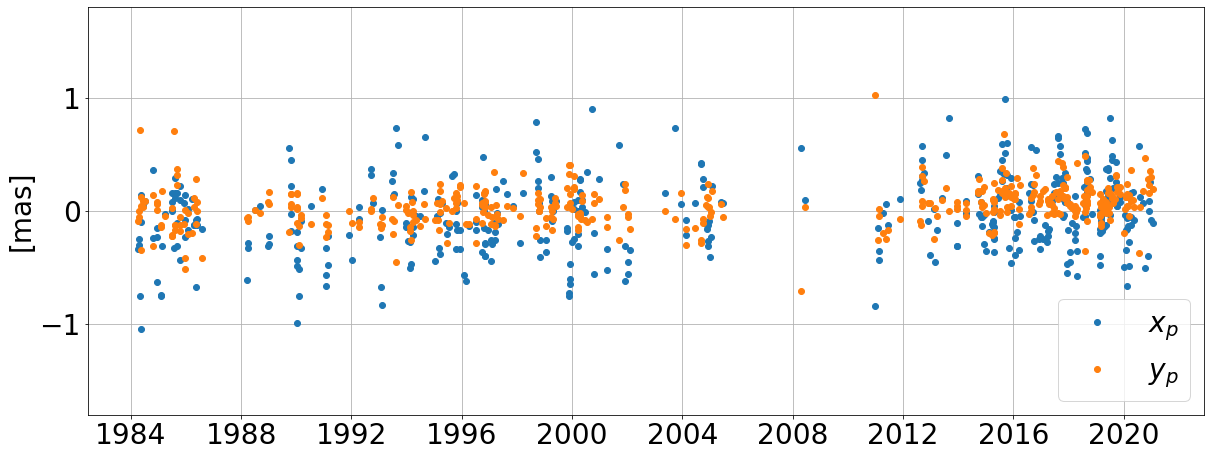}
    \caption{Estimated values of $x_p$ and $y_p$ for NTSL solution from IMLS subtracted from the estimated values of the standard solution for the subset All15.}\label{fig:ntsl_all15all}
\end{figure}

As the station coordinates and PMC are also correlated, we checked the effect when adding \ac{NTL} for the estimation of PMC from LUNAR, as we did for $\Delta$UT estimation. The \ac{NTL} was added only from IMLS, in three individual loading components: NTAL, NTOL, and HYDL, and a fourth component of their combination: NTSL. The effect of \ac{NTL}, as expected, is of similar - small - magnitude on the estimated PMC values from all subsets. In Fig. \ref{fig:ntsl_all15all}, we show the effect when adding NTSL to the standard solution for the estimation of PMC from the subset All15. Additionally, in Table \ref{tab:ntsl_xpyp_acc}, we show the mean values of the accuracy obtained for the estimation of PMC from the subsets All10 and All15 for the standard solution and the \ac{NTL} solutions from IMLS.


\begin{table}
\caption{Mean values of accuracy of the PMC values for the standard and the \ac{NTL} solutions  for the subsets All10 and All15.}\label{tab:ntsl_xpyp_acc}
\centering
\begin{tabular}{ccP{0.8cm}P{0.8cm}P{1.0cm}P{1.0cm}}
\hline\noalign{\smallskip}
&Loading &\multicolumn{2}{c}{All10 [mas]} &\multicolumn{2}{c}{All15 [mas]} \\
& &xp &yp &xp &yp \\
\noalign{\smallskip}\hline\noalign{\smallskip}
Std & &5.37 &5.18 &4.36 &4.40 \\
\noalign{\smallskip}\hline\noalign{\smallskip}
\multirow{4}[6]{*}{IMLS} &NTAL &5.34 &5.15 &4.34 &4.37\\
\noalign{\smallskip}\cline{2-6}\noalign{\smallskip}
&NTOL &5.35 &5.17 &4.35 &4.38\\
\noalign{\smallskip}\cline{2-6}\noalign{\smallskip}
&HYDL &5.35 &5.16 &4.35 &4.39\\
\noalign{\smallskip}\cline{2-6}\noalign{\smallskip}
&NTSL &5.31 &5.13 &4.32 &4.35\\
\noalign{\smallskip}\hline
\noalign{\smallskip}
\end{tabular}
\end{table}

When adding \ac{NTL} from any loadings, there was no change observed in the correlations of the PMC with the non-ERP parameters of LUNAR. In our previous study \citep{singh_etal_2021}, we discussed the benefit of reduction of the annual signal in LLR residuals when adding \ac{NTL}. However, as the data from the subsets is temporally extremely unevenly distributed, any effect the addition of \ac{NTL} might have at the annual signal is masked by noise. 

\begin{figure}[!ht]
    \centering
    \includegraphics[width=0.49\textwidth]{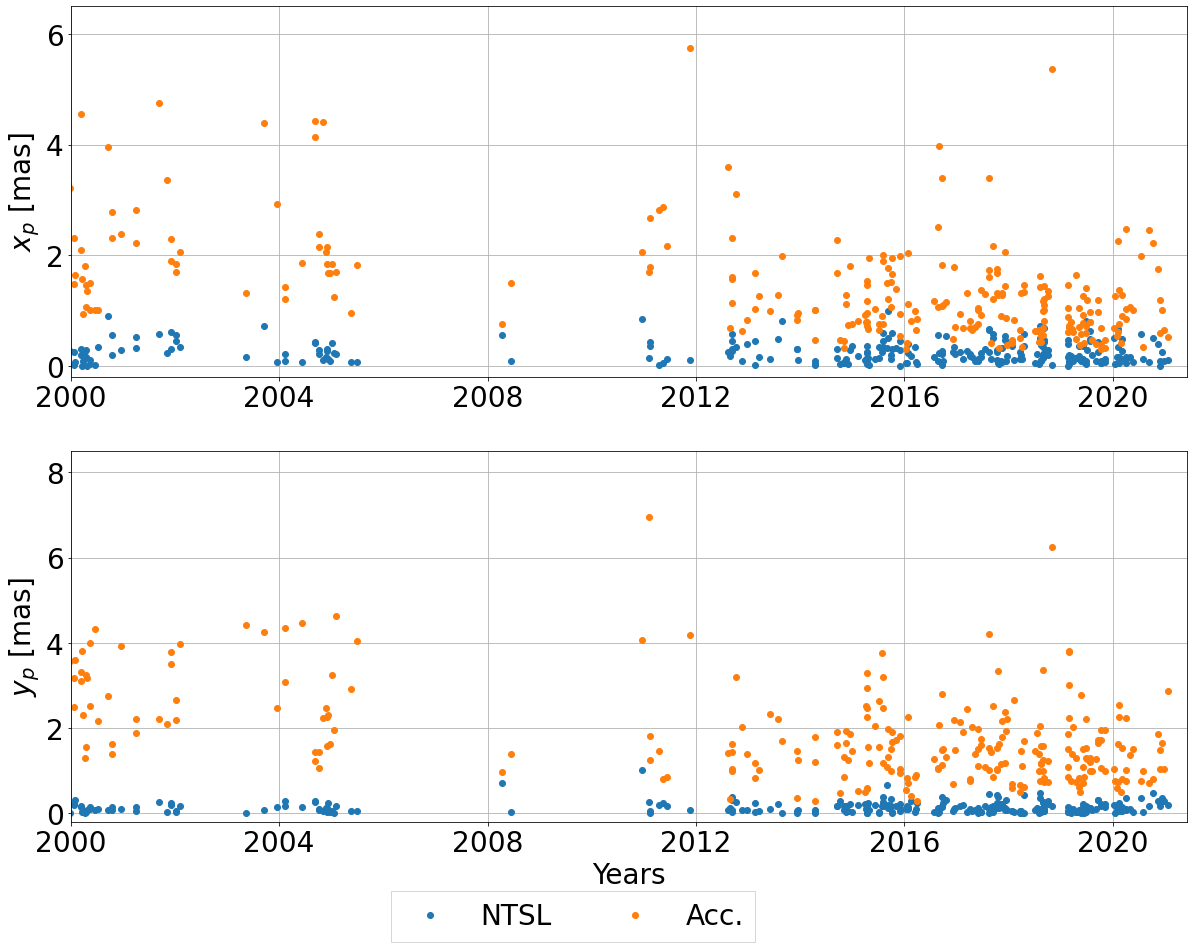}
    \caption{Absolute values of the effect of addition of NTSL on $x_p$ and $y_p$ and the values of accuracies for the estimated $x_p$ and $y_p$ for the subset All15, after 2000.0.}\label{fig:ntsl_all15all_comp}
\end{figure}

In Fig. \ref{fig:ntsl_all15all_comp}, we show a comparison of the effect of NTSL on the PMC (absolute values) with the accuracies of the estimation of $x_p$ and $y_p$ values for the results after 2000.0 for the subset All15, as the best accuracies are obtained in this time span. It can be seen that the absolute effect of NTSL is significantly smaller than the accuracies of the $x_p$ and $y_p$ values. In spite of the effect being smaller, there is a small improvement in the overall accuracy of the estimated $\Delta$UT values when including the NTSL, see Table \ref{tab:ntsl_xpyp_acc}. Overall, the accuracy of the PMC improves for the IMLS NTSL solutions (about \SI{1}{\percent} for both components of both subsets). As mentioned previously, these changes are in sync with the findings from our previous study \citep{singh_etal_2021}.


\section{Conclusions}\label{sec:conclusions}
In this study, we estimated the Earth rotation parameters, Earth rotation phase, $\Delta$UT, and the terrestrial pole offsets, $x_p$ and $y_p$, from LLR analysis for eight different subsets of nights in the LLR time span. We estimated the \ac{ERP} by performing a least square adjustment using the Gauss-Markov model. For the analysis, we kept the velocities of the LLR observatories fixed to the ITRF2014 solution values, and estimated $\Delta$UT and the \ac{PMC} separately, to avoid high correlations between them. For the estimation of \ac{PMC}, we estimated $x_p$ and $y_p$ together and separately. We discussed the differences of the estimated \ac{ERP} results to the IERS 14 C04 series, for the different subsets, analysed the accuracy of their estimation, and additionally discussed the effect of addition of \ac{NTL} on the ERP estimation. 

Generally, the estimation of \ac{ERP} shows a significant improvement over the time span of the subsets considered in this study, with the mean accuracy of the estimated values reducing to (almost) a third of its value when comparing the results split at 0h UTC 01.01.2000. The best possible resolution of estimated ERP (on Earth's surface) is of \SI{7.3}{\milli\meter} (i.e. \SI{15.89}{\micro\second}) from $\Delta$UT estimation, and of about \SI{4}{cm} from the estimation of PMC (i.e. \SI{1.16}{\mas} for $x_p$ and \SI{1.52}{\mas} for $y_p$). Compared to other space geodetic techniques, the results from LLR still lag behind, however are still important, as LLR provides the longest time series of any space geodetic technique and is the only technique other than \ac{VLBI} which provides $\Delta$UT values. With the more accurate and more frequent LLR data in future, possibly with the measurements using differential LLR \citep{Viswanathan2021}, better results for ERP can be expected from LLR.

LLR is more sensitive to the estimation of $\Delta$UT values than PMC values. This is due to low number of \acp{NP} per night used within the estimation, combined with the fact that changes per night are larger for $\Delta$UT than for the PMC. Furthermore, due to the distribution of the LLR data (most NPs are measured from OCA), the estimation of PMC from LLR is more sensitive to x-direction than to y-direction. The values of the mean accuracy obtained from different subsets, shown in Table \ref{tab:pol_subsets}, where the values of $x_p$ are smaller than $y_p$, and the improvement in the values of $x_p$ is more significant than in $y_p$, when comparing the results from the old NPs (before 2000.0) to the results from the new NPs (after 2000.0).

For the estimation of PMC, with the estimation of $x_p$ and $y_p$ together and separately, we were able to assess the stability of the calculation. With the subsets which were selected with a strict selection criteria of either 10 or 15 NPs per night, and when the number of estimated nights was not too few (such as for the subset Apollo10, 63 nights) the results of the estimated $x_p$ and $y_p$ values were not too different from each other when estimated individually or together.

The estimated ERP are primarily correlated to the positions of the observatory the nights were selected from. With a more strict selection criteria (15 NPs per night instead of 5 NPs per night), the correlations reduce significantly, however, this comes at the cost of having fewer nights at which the ERP can be estimated. The estimated PMC (when estimated together) are highly correlated to each other ($x_p$ and the $y_p$ values of the same night), with correlations going as high as \SI{100}{\percent}. These correlations come down to a highest of \SI{80}{\percent} with the strict selection criteria of 15NPs per night from one observatory, however are still too high. With the less strict selection criteria of 5 NPs per night, the ERP show correlations up to \SI{20}{\percent} and \SI{30}{\percent} with some parameters such as reflectors, the rotational time lags of the Earth (diurnal and semi-diurnal, due to the tidal effect of the Moon), etc., which disappear (or become smaller, and therefore irrelevant) when implementing the stricter selection criteria. 
 
The \ac{NTL} was applied as observation level corrections in our LLR analysis and added as three different loading constituents for mass redistribution in atmosphere, oceans, and land water from the IMLS dataset. The impact when adding all three components of \ac{NTL} on \ac{ERP} estimation is smaller in magnitude than the accuracies of the ERP values, for both $\Delta$UT and PMC. However, it leads to about \SI{1}{\percent} improvement in the accuracies obtained, and is therefore recommended to be added for ERP estimation.

\section*{Acknowledgements}
We acknowledge with thanks that the processed LLR data, since 1969, has been obtained under the efforts of the personnel at the Observatoire de la C\^{o}te d'Azur in France, the LURE Observatory in Maui, Hawaii, the McDonald Observatory in Texas, the Apache Point Observatory in New Mexico, the Matera Laser Ranging observatory in Italy, and the Wettzell Laser Ranging System in Germany. This research was funded by the German Aerospace Center's (DLR) Institute for Satellite Geodesy and Inertial Sensing, and Deutsche Forschungsgemeinschaft (DFG, German Research Foundation) under Germany’s Excellence Strategy EXC 2123 QuantumFrontiers, Project-ID 390837967. We would additionally like to thank Franz Hofmann for his contributions to LUNAR.

\section*{Data availability}
LLR data is collected, archived, and distributed under the auspices of the International Laser Ranging Service (ILRS) \citep{ilrs2019}; and downloaded from the website\footnote{\url{https://cddis.nasa.gov/About/CDDIS_File_Download_Documentation.html}}. The IMLS non-tidal loading dataset is freely available on the IMLS website\footnote{\url{http://massloading.net/}}; and downloaded for the LLR observatories from the pre-computed time series of 1272 space geodesy sites. The KEOF COMB2019 EOP time series is freely available at the website \footnote{\url{https://keof.jpl.nasa.gov/combinations/latest/}}, and the IERS 14 C04 EOP time series is freely available at the website \footnote{\url{https://www.iers.org/IERS/EN/DataProducts/EarthOrientationData/eop.html;jsessionid=4593E0A857CFBA2E4141374D13DA2F55.live2}}.
\section*{Author contributions}
All authors contributed to the development of this study and provided ideas to its content. Data collection and analysis were performed by VVS, LB, and MZ. The first draft of the manuscript was written by VVS, and all authors commented on previous versions of the manuscript. All authors read and approved the final manuscript. 


\bibliography{main}

\begin{thebibliography}{46}
\providecommand{\natexlab}[1]{#1}
\providecommand{\url}[1]{\texttt{#1}}
\expandafter\ifx\csname urlstyle\endcsname\relax
  \providecommand{\doi}[1]{doi: #1}\else
  \providecommand{\doi}{doi: \begingroup \urlstyle{rm}\Url}\fi

\bibitem[Bauer(1989)]{Bauer1989}
R.~Bauer.
\newblock \emph{{Bestimmung von Parametern des Erde-Mond-Systems - Ein Beitrag
  zur Modellerweiterung und Bewertung, Ergebnisse -}}.
\newblock PhD thesis, Technische Universit\"at M\"unchen, 1989.
\newblock {Deutsche Geod\"atische Kommission bei der Bayerischen Akademie der
  Wissenschaften, Reihe C, Nr. 353}.

\bibitem[Biskupek(2015)]{bisk2015}
L.~Biskupek.
\newblock \emph{{Bestimmung der Erdorientierung mit Lunar Laser Ranging}}.
\newblock PhD thesis, Leibniz University Hannover, Deutsche Geodätische
  Kommission bei der Bayerischen Akademie der Wissenschaften, Reihe C, Nr. 742,
  2015.
\newblock doi: 10.15488/4721.

\bibitem[Biskupek et~al.(2021)Biskupek, M\"uller, and Torre]{biskupek21}
L.~Biskupek, J.~M\"uller, and J.-M. Torre.
\newblock Benefit of new high-precision llr data for the determination of
  relativistic parameters.
\newblock \emph{Universe}, 7\penalty0 (2), 2021.
\newblock \doi{10.3390/universe7020034}.

\bibitem[Bizouard et~al.(2019)Bizouard, Lambert, Gattano, Becker, and
  Richard]{Bizouard2019}
C.~Bizouard, S.~Lambert, C.~Gattano, O.~Becker, and J.-Y. Richard.
\newblock {The IERS EOP 14C04 solution for Earth orientation parameters
  consistent with ITRF 2014}.
\newblock \emph{Journal of Geodesy}, 93\penalty0 (5):\penalty0 621--633, 2019.
\newblock \doi{10.1007/s00190-018-1186-3}.

\bibitem[Capitaine(2017)]{capitain_eop17}
N.~Capitaine.
\newblock {The Determination of Earth Orientation by VLBI and GNSS: Principles
  and Results}.
\newblock In E.~F. Arias, L.~Combrinck, P.~Gabor, C.~Hohenkerk, and P.~K.
  Seidelmann, editors, \emph{The Science of Time 2016}, pages 167--196, Cham,
  2017. Springer International Publishing.

\bibitem[Chabé et~al.(2020)Chabé, Courde, Torre, Bouquillon, Bourgoin, Aimar,
  Albanèse, Chauvineau, Mariey, Martinot-Lagarde, Maurice, Phung, Samain, and
  Viot]{Chabe2020}
J.~Chabé, C.~Courde, J.-M. Torre, S.~Bouquillon, A.~Bourgoin, M.~Aimar,
  D.~Albanèse, B.~Chauvineau, H.~Mariey, G.~Martinot-Lagarde, N.~Maurice,
  D.-H. Phung, E.~Samain, and H.~Viot.
\newblock {Recent Progress in Lunar Laser Ranging at Grasse Laser Ranging
  Station}.
\newblock \emph{Earth and Space Science}, 7:\penalty0 e2019EA000785, 2020.
\newblock \doi{10.1029/2019EA000785}.

\bibitem[Combrinck({2009})]{Combrinck09}
W.~Combrinck.
\newblock {Products of Space Geodesy and Links to Earth Science and Astronomy}.
\newblock {2009}.
\newblock ISSN {2214-4609}.
\newblock \doi{10.3997/2214-4609-pdb.241.combrinck_wl_paper1}.
\newblock {Conference Proceedings, 11th SAGA Biennial Technical Meeting and
  Exhibition, cp-241-00039 }.

\bibitem[Dickey et~al.(1985)Dickey, Newhall, and Williams]{dickeyetal1985}
J.~O. Dickey, X.~X. Newhall, and J.~G. Williams.
\newblock {Earth Orientation From Lunar Laser Ranging and an Error Analysis of
  Polar Motion Services}.
\newblock \emph{{Journal of Geophysical Research}}, 90, 10 1985.
\newblock \doi{10.1029/JB090iB11p09353}.

\bibitem[Dill and Dobslaw(2013)]{gfz_ntsl}
R.~Dill and H.~Dobslaw.
\newblock Numerical simulations of global-scale high-resolution hydrological
  crustal deformations.
\newblock \emph{Journal of Geophysical Research: Solid Earth}, 118\penalty0
  (9):\penalty0 5008--5017, 2013.
\newblock \doi{10.1002/jgrb.50353}.

\bibitem[Egger(1985)]{Egger1985}
D.~Egger.
\newblock \emph{{Systemanalyse der Laserentfernungsmessung}}.
\newblock PhD thesis, Technische Universität München, 1985.
\newblock {Deutsche Geod\"atische Kommission bei der Bayerischen Akademie der
  Wissenschaften, Reihe C, Nr. 311}.

\bibitem[Farrell(1972)]{farrell72}
W.~E. Farrell.
\newblock {Deformation of the Earth by surface loads}.
\newblock \emph{Rev. Geophys. and Spac. Phys.}, 10(3):\penalty0 751–797,
  1972.
\newblock \doi{10.1029/RG010i003p00761}.

\bibitem[Folkner et~al.(2014)Folkner, Williams, Boggs, Park, and
  Kuchynka]{jplde}
W.~M. Folkner, J.~G. Williams, D.~H. Boggs, R.~S. Park, and P.~Kuchynka.
\newblock {The Planetary and Lunar Ephemerides DE430 and DE431}.
\newblock \emph{IPN Progress Report 42-196}, 2014.

\bibitem[Gelaro et~al.(2017)Gelaro, McCarty, Suárez, Todling, Molod, Takacs,
  Randles, Darmenov, Bosilovich, Reichle, Wargan, Coy, Cullather, Draper,
  Akella, Buchard, Conaty, da~Silva, Gu, Kim, Koster, Lucchesi, Merkova,
  Nielsen, Partyka, Pawson, Putman, Rienecker, Schubert, Sienkiewicz, and
  Zhao]{merra2}
R.~Gelaro, W.~McCarty, M.~J. Suárez, R.~Todling, A.~Molod, L.~Takacs, C.~A.
  Randles, A.~Darmenov, M.~G. Bosilovich, R.~Reichle, K.~Wargan, L.~Coy,
  R.~Cullather, C.~Draper, S.~Akella, V.~Buchard, A.~Conaty, A.~M. da~Silva,
  W.~Gu, G.-K. Kim, R.~Koster, R.~Lucchesi, D.~Merkova, J.~E. Nielsen,
  G.~Partyka, S.~Pawson, W.~Putman, M.~Rienecker, S.~D. Schubert,
  M.~Sienkiewicz, and B.~Zhao.
\newblock {The Modern-Era Retrospective Analysis for Research and Applications,
  Version 2 (MERRA-2)}.
\newblock \emph{Journal of Climate - American Meteorological Society},
  30\penalty0 (14):\penalty0 5419–5454, 2017.
\newblock \doi{10.1175/JCLI-D-16-0758.1}.

\bibitem[Gleixner(1986)]{Gleixner1986}
H.~Gleixner.
\newblock \emph{{Ein Beitrag zur Ephemeridenrechnung und Parametersch\"atzung
  im Erde-Mond-System}}.
\newblock PhD thesis, Technische Universit\"at M\"unchen, 1986.
\newblock {Deutsche Geod\"atische Kommission bei der Bayerischen Akademie der
  Wissenschaften, Reihe C, Nr. 319}.

\bibitem[Glomsda et~al.(2020)Glomsda, Bloßfeld, Seitz, and
  Seitz]{glomsda_etal2020}
M.~Glomsda, M.~Bloßfeld, M.~Seitz, and F.~Seitz.
\newblock {Benefits of non-tidal loading applied at distinct levels in VLBI
  analysis}.
\newblock \emph{Journal of Geodesy}, 94\penalty0 (90):\penalty0 1--19, 2020.
\newblock \doi{10.1007/s00190-020-01418-z}.

\bibitem[Hersbach et~al.(2018)Hersbach, de~Rosnay, Bell, Schepers, Simmons,
  Soci, Abdalla, Balmaseda, Balsamo, Bechtold, Berrisford, Bidlot,
  de~Boisséson, Bonavita, Browne, Buizza, Dahlgren, Dee, Dragani, Diamantakis,
  Flemming, Forbes, Geer, Haiden, Hólm, Haimberger, Hogan, Horányi,
  Janisková, Laloyaux, Lopez, Muñoz-Sabater, Peubey, Radu, Richardson,
  Thépaut, Vitart, Yang, Zsótér, and Zuo]{era5}
H.~Hersbach, P.~de~Rosnay, B.~Bell, D.~Schepers, A.~Simmons, C.~Soci,
  S.~Abdalla, M.~A. Balmaseda, G.~Balsamo, P.~Bechtold, P.~Berrisford,
  J.~Bidlot, E.~de~Boisséson, M.~Bonavita, P.~Browne, R.~Buizza, P.~Dahlgren,
  D.~Dee, R.~Dragani, M.~Diamantakis, J.~Flemming, R.~Forbes, A.~Geer,
  T.~Haiden, E.~Hólm, L.~Haimberger, R.~Hogan, A.~Horányi, M.~Janisková,
  P.~Laloyaux, P.~Lopez, J.~Muñoz-Sabater, C.~Peubey, R.~Radu, D.~Richardson,
  J.-N. Thépaut, F.~Vitart, X.~Yang, E.~Zsótér, and H.~Zuo.
\newblock {Operational global reanalysis: progress, future directions and
  synergies with NWP}.
\newblock Technical Report~27, European Centre for Medium Range Weather
  Forecasts, Shinfield Park, Reading, Berkshire RG2 9AX, England, 2018.

\bibitem[Hofmann(2017)]{hofmann17}
F.~Hofmann.
\newblock \emph{{Lunar Laser Ranging –verbesserte Modellierung der
  Monddynamikund Schätzung relativistischer Parameter}}.
\newblock PhD thesis, Leibniz University Hannover, Deutsche Geodätische
  Kommission bei der Bayerischen Akademie der Wissenschaften, Reihe C, Nr. 797,
  2017.

\bibitem[Hofmann and M\"uller(2018)]{hof_mu18}
F.~Hofmann and J.~M\"uller.
\newblock {Relativistic Tests with Lunar Laser Ranging}.
\newblock \emph{Classical and Quantum Gravity}, 35-035015\penalty0 (3), 2018.
\newblock \doi{10.1088/1361-6382/aa8f7a}.

\bibitem[Hofmann et~al.(2018)Hofmann, Biskupek, and M\"uller]{hof_etal18}
F.~Hofmann, L.~Biskupek, and J.~M\"uller.
\newblock {Contributions to reference systems from Lunar Laser Ranging using
  the IfE analysis model}.
\newblock \emph{Journal of Geodesy}, 92:\penalty0 975–987, 2018.
\newblock \doi{10.1007/s00190-018-1109-3}.

\bibitem[Jungclaus et~al.(2013)Jungclaus, Fischer, Haak, Lohmann, Marotzke,
  Matei, Mikolajewicz, Notz, and von Storch]{mpiom}
J.~H. Jungclaus, N.~Fischer, H.~Haak, K.~Lohmann, J.~Marotzke, D.~Matei,
  U.~Mikolajewicz, D.~Notz, and J.~S. von Storch.
\newblock {Characteristics of the ocean simulations in the Max Planck Institute
  Ocean Model (MPIOM) the ocean component of the MPI‐Earth system model}.
\newblock \emph{Journal of Advances in Modeling Earth Systems}, 5\penalty0
  (2):\penalty0 422– 446, 2013.
\newblock \doi{10.1002/jame.20023}.

\bibitem[Kopeikin et~al.(2008)Kopeikin, Pavlis, Pavlis, Brumberg, Escapa,
  Getino, Gusev, Müller, Ni, and Petrova]{kopeikin2008}
S.~Kopeikin, E.~Pavlis, D.~Pavlis, V.~Brumberg, A.~Escapa, J.~Getino, A.~Gusev,
  J.~Müller, W.-T. Ni, and N.~Petrova.
\newblock {Prospects in the orbital and rotational dynamics of the Moon with
  the advent of sub-centimeter lunar laser ranging}.
\newblock \emph{{Advances in Space Research}}, 42\penalty0 (8):\penalty0
  1378--1390, 2008.
\newblock ISSN 0273-1177.
\newblock \doi{10.1016/j.asr.2008.02.014}.

\bibitem[M\"uller(1991)]{Muller1991}
J.~M\"uller.
\newblock \emph{{Analyse von Lasermessungen zum Mond im Rahmen einer
  post-Newton'schen Theorie}}.
\newblock PhD thesis, Technische Universit\"at M\"unchen, 1991.
\newblock {Deutsche Geod\"atische Kommission bei der Bayerischen Akademie der
  Wissenschaften, Reihe C, Nr. 383}.

\bibitem[M\"uller et~al.(2009)M\"uller, Biskupek, Oberst, and
  Schreiber]{mueller2009}
J.~M\"uller, L.~Biskupek, J.~Oberst, and U.~Schreiber.
\newblock {Contribution of Lunar Laser Ranging to Realise Geodetic Reference
  Systems}.
\newblock In \emph{Geodetic Reference Frames. International Association of
  Geodesy Symposia}, volume 134, pages 55--59. Springer Berlin Heidelberg,
  Berlin, Heidelberg, 10 2009.
\newblock \doi{10.1007/978-3-642-00860-3-8}.

\bibitem[M\"uller et~al.(2012)M\"uller, Hofmann, and Biskupek]{mueller_etal12}
J.~M\"uller, F.~Hofmann, and L.~Biskupek.
\newblock {Testing various facets of the equivalence principle using lunar
  laser ranging}.
\newblock \emph{Classical and Quantum Gravity}, 29:\penalty0 184006, 09 2012.
\newblock \doi{10.1088/0264-9381/29/18/184006}.

\bibitem[M\"uller et~al.(2014)M\"uller, Biskupek, Hofmann, and
  Mai]{Muller2014a}
J.~M\"uller, L.~Biskupek, F.~Hofmann, and E.~Mai.
\newblock {Lunar laser ranging and relativity}.
\newblock In S.~M. Kopeikin, editor, \emph{{Frontiers in relativistic celestial
  Mechanics}}, volume 2: Applications and Experiments, pages 103--156. Walter
  de Gruyter, Berlin, 2014.

\bibitem[M\"uller et~al.(2019)M\"uller, Murphy, Schreiber, Shelus, Torre,
  Williams, Boggs, Bouquillon, Bourgoin, and Hofmann]{mueller19}
J.~M\"uller, T.~W. Murphy, U.~Schreiber, P.~J. Shelus, J.~M. Torre, J.~G.
  Williams, D.~H. Boggs, S.~Bouquillon, A.~Bourgoin, and F.~Hofmann.
\newblock {Lunar Laser Ranging: a tool for general relativity, lunar geophysics
  and Earth science}.
\newblock \emph{Journal of Geodesy}, 93:\penalty0 2195–2210, 2019.
\newblock \doi{10.1007/s00190-019-01296-0}.

\bibitem[Murphy(2013)]{murp13}
T.~W. Murphy.
\newblock {Lunar laser ranging: the millimeter challenge}.
\newblock \emph{Reports on Progress in Physics}, 76:\penalty0 076901, 2013.
\newblock \doi{10.1088/0034-4885/76/7/076901}.

\bibitem[Pavlov et~al.(2016)Pavlov, Williams, and Suvorkin]{pav16}
D.~A. Pavlov, J.~G. Williams, and V.~V. Suvorkin.
\newblock {Determining parameters of Moon’s orbital and rotational motion
  from LLR observations using GRAIL and IERS-recommended models}.
\newblock \emph{Celestial Mechanics and Dynamical Astronomy}, 126:\penalty0
  61--88, 2016.
\newblock \doi{10.1007/s10569-016-9712-1}.

\bibitem[{Pearlman} et~al.(2019){Pearlman}, {Noll}, {Pavlis}, {Lemoine},
  {Combrink}, {Degnan}, {Kirchner}, and {Schreiber}]{ilrs2019}
M.~R. {Pearlman}, C.~E. {Noll}, E.~C. {Pavlis}, F.~G. {Lemoine}, L.~{Combrink},
  J.~J. {Degnan}, G.~{Kirchner}, and U.~{Schreiber}.
\newblock {The ILRS: approaching 20 years and planning for the future}.
\newblock \emph{Journal of Geodesy}, 93\penalty0 (11):\penalty0 2161--2180,
  Nov. 2019.
\newblock \doi{10.1007/s00190-019-01241-1}.

\bibitem[Petit and Luzum(2010)]{Petit2010}
G.~Petit and B.~Luzum, editors.
\newblock \emph{{IERS Conventions 2010}}.
\newblock Number~36 in IERS Technical Note. Verlag des Bundesamtes f\"ur
  Kartographie und Geod\"asie, Frankfurt am Main, 2010.

\bibitem[Petrov(2015)]{imls_ntsl}
L.~Petrov.
\newblock {The International Mass Loading Service}, 2015.
\newblock {http://arxiv.org/abs/1503.00191}.

\bibitem[Petrov and Boy(2004)]{petrov_boy04}
L.~Petrov and J.-P. Boy.
\newblock Study of the atmospheric pressure loading signal in very long
  baseline interferometry observations.
\newblock \emph{Journal of Geophysical Research: Solid Earth}, 109\penalty0
  (B3), 2004.
\newblock ISSN 148-227.
\newblock \doi{10.1029/2003JB002500}.

\bibitem[Ratcliff and Gross(2020)]{Ratcliff2020}
J.~T. Ratcliff and R.~S. Gross.
\newblock {Combinations of Earth OrientationMeasurements: SPACE2019,
  COMB2019,and POLE2019}.
\newblock Technical Report JPL Publication 20-3, Jet Propulsion Laboratory,
  2020.

\bibitem[Schuh and Behrend(2012)]{SCHUH201268}
H.~Schuh and D.~Behrend.
\newblock {VLBI: A fascinating technique for geodesy and astrometry}.
\newblock \emph{Journal of Geodynamics}, 61:\penalty0 68--80, 2012.
\newblock ISSN 0264-3707.
\newblock \doi{10.1016/j.jog.2012.07.007}.

\bibitem[Sciarretta et~al.(2010)Sciarretta, Luceri, Pavlis, and
  Bianco]{Sciarretta2010}
C.~Sciarretta, V.~Luceri, E.~Pavlis, and G.~Bianco.
\newblock {The ILRS EOP time series}.
\newblock \emph{Artificial Satellites}, 45:\penalty0 41--48, 01 2010.
\newblock \doi{10.2478/v10018-010-0004-9}.

\bibitem[Singh et~al.(2021)Singh, Biskupek, M\"uller, and
  Zhang]{singh_etal_2021}
V.~V. Singh, L.~Biskupek, J.~M\"uller, and M.~Zhang.
\newblock {Impact of non-tidal station loading in LLR}.
\newblock \emph{Advances in Space Research}, 67\penalty0 (12):\penalty0
  3925--3941, 2021.
\newblock ISSN 0273-1177.
\newblock \doi{10.1016/j.asr.2021.03.018}.

\bibitem[Sun(2017)]{sun_yu_gcm}
Y.~Sun.
\newblock \emph{{Estimating geocenter motion and changes in the Earth’s
  dynamic oblateness from GRACE and geophysical models}}.
\newblock PhD thesis, TU Delft Physical and Space Geodesy, 2017.
\newblock {doi: 10.4233/uuid:7fe64dde-7fb5-4392-8160-da6f7916dc6b}.

\bibitem[VanderPlas(2017)]{ls_ana}
J.~T. VanderPlas.
\newblock {Understanding the Lomb-Scargle Periodogram}.
\newblock \emph{The Astrophysical Journal Supplement Series}, 236\penalty0
  (1):\penalty0 16, 2017.
\newblock \doi{10.3847/1538-4365/aab766}.

\bibitem[Viswanathan et~al.(2019)Viswanathan, Rambaux, andJ. Laskar, and
  Gastineau]{viswa19}
V.~Viswanathan, N.~Rambaux, A.~F. andJ. Laskar, and M.~Gastineau.
\newblock {Observational Constraint on the Radius and Oblateness of the Lunar
  Core-Mantle Boundary}.
\newblock \emph{Geophysical Research Letters}, 46:\penalty0 7295–7303, 2019.
\newblock \doi{10.1029/2019GL082677}.

\bibitem[Viswanathan et~al.(2021)Viswanathan, Mazarico, Merkowitz, Williams,
  Turyshev, Currie, Ermakov, Rambaux, Fienga, Courde, Chabé, Torre, Bourgoin,
  Schreiber, Eubanks, Wu, Dequal, Dell'Agnello, Biskupek, M\"uller, and
  Kopeikin]{Viswanathan2021}
V.~Viswanathan, E.~Mazarico, S.~Merkowitz, J.~Williams, S.~Turyshev, D.~Currie,
  A.~Ermakov, N.~Rambaux, A.~Fienga, C.~Courde, J.~Chabé, J.-M. Torre,
  A.~Bourgoin, K.~Schreiber, M.~Eubanks, C.~Wu, D.~Dequal, S.~Dell'Agnello,
  L.~Biskupek, J.~M\"uller, and S.~Kopeikin.
\newblock {Extending Science from Lunar Laser Ranging}.
\newblock \emph{Bulletin of the American Astronomical Society}, 53, 03 2021.
\newblock \doi{10.3847/25c2cfeb.3dc2e5e4}.

\bibitem[Williams et~al.(2006)Williams, Turyshev, Boggs, and Ratcliff]{will06}
J.~G. Williams, S.~G. Turyshev, D.~H. Boggs, and J.~T. Ratcliff.
\newblock {Lunar laser ranging science: Gravitational physics and lunar
  interior and geodesy}.
\newblock \emph{Advances in Space Research}, 37:\penalty0 67--71, 2006.
\newblock \doi{10.1016/j.asr.2005.05.013}.

\bibitem[Williams et~al.(2009)Williams, Turyshev, and Boggs]{Williams2009b}
J.~G. Williams, S.~G. Turyshev, and D.~H. Boggs.
\newblock {Lunar Laser Ranging Tests of the Equivalence Principle with the
  Earth and Moon}.
\newblock \emph{International Journal of Modern Physics D}, 18\penalty0
  (7):\penalty0 1129--1175, 2009.
\newblock \doi{10.1142/S021827180901500X}.

\bibitem[Williams et~al.(2013)Williams, Boggs, and Folkner]{de430_other}
J.~G. Williams, D.~H. Boggs, and W.~M. Folkner.
\newblock {DE430 Lunar Orbit, Physical Librations, and Surface Coordinates}.
\newblock Technical report, Jet Propulsion Laboratory, California Institute of
  Technology, Pasadena, CA, USA, 7 2013.
\newblock Interoffice memorandum, IOM 335-JW,DB,WF-20130722-016.

\bibitem[Zajdel et~al.(2020)Zajdel, Sośnica, Bury, Dach, and
  Prange]{Zajdel2020}
R.~Zajdel, K.~Sośnica, G.~Bury, R.~Dach, and L.~Prange.
\newblock System-specific systematic errors in earth rotation parameters
  derived from gps, glonass, and galileo.
\newblock \emph{GPS Solutions}, 24, 05 2020.
\newblock \doi{10.1007/s10291-020-00989-w}.

\bibitem[Zerhouni and Capitaine(2009)]{zc09}
W.~Zerhouni and N.~Capitaine.
\newblock {Celestial pole offsets from lunar laser ranging and comparison with
  VLBI}.
\newblock \emph{{Astronomy $\&$ Astrophysics}}, 507:\penalty0 1687--1695, 12
  2009.
\newblock \doi{10.1051/0004-6361/200912644}.

\bibitem[Zhang et~al.(2020)Zhang, M{\"{u}}ller, and Biskupek]{Zhang2020}
M.~Zhang, J.~M{\"{u}}ller, and L.~Biskupek.
\newblock {Test of the equivalence principle for galaxy's dark matter by lunar
  laser ranging}.
\newblock \emph{Celestial Mechanics and Dynamical Astronomy}, 132\penalty0
  (4):\penalty0 25, 2020.
\newblock ISSN 1572-9478.
\newblock \doi{10.1007/s10569-020-09964-6}.

\end{thebibliography}




\acrodef{LLR}{Lunar Laser Ranging}
\acrodef{NTSL}{non-tidal station loading}
\acrodef{GFZ}{German Research Centre for Geosciences}
\acrodef{IMLS}{International Mass Loading Service}
\acrodef{EOST}{loading service of University of Strasbourg}
\acrodef{NP}{normal point}
\acrodef{OCA}{Côte d'Azur Observatory, France}
\acrodef{WLRS}{Geodetic Observatory Wettzell, Germany}
\acrodef{ERP}{Earth Rotation Parameter}
\acrodef{EOP}{Earth Orientation Parameter}
\acrodef{APOLLO}{Apache Point Observatory Lunar Laser ranging Operation, USA}
\acrodef{LURE}{Lure Observatory on Maui island, Hawaii, USA}
\acrodef{MLRO}{Matera Laser Ranging Observatory, Italy}
\acrodef{MLRS}{McDonald Laser Ranging Station, USA}
\acrodef{LUNAR}[LUNAR]{LUNar laser ranging Analysis softwaRe}
\acrodef{IERS}{International Earth Rotation and Reference Systems Service}
\acrodef{LOD}{Length-of-Day}
\acrodef{GNSS}{Global Navigation Satellite System}
\acrodef{SLR}{Satellite Laser Ranging}
\acrodef{VLBI}{Very Long Baseline Interferometry}
\acrodef{DORIS}{Doppler Orbitography and Radiopositioning Integrated by Satellite}
\acrodef{PMC}{polar motion coordinates}
\acrodef{GCRS}{Geocentric Celestial Reference System}
\acrodef{ITRS}{International Terrestrial Reference System}
\acrodef{NTL}{non-tidal loading}
\acrodef{ICRS}{International Celestial Reference System}

\end{document}